\documentclass[journal=jctcce,manuscript=article]{achemso}

\usepackage{amssymb}
\usepackage{amsmath}
\usepackage{graphicx}
\usepackage{mathtools}
\usepackage{xcolor}
\usepackage{hyperref}
\usepackage{placeins}
\usepackage{siunitx}
\usepackage{subcaption}
\usepackage{cleveref}
\usepackage[utf8]{inputenc}

\makeatletter

\newcommand{\aims}{\texttt{FHI-aims}\cite{Sinstein,aims-paper,aims-Havu2009,aims-Yu2018,aims-Ren2012,aims-Ihrig2015} }

\newcommand\change[1]{#1}
\newcommand\note[1]{}

\author{Jakob Filser}
\email{jakob.filser@tum.de}
\affiliation{Chair for Theoretical Chemistry and Catalysis Research Center, Technische Universit\"at M\"unchen, Lichtenbergstr. 4, 85747 Garching, Germany}
\author{Karsten Reuter}
\affiliation{Chair for Theoretical Chemistry and Catalysis Research Center, Technische Universit\"at M\"unchen, Lichtenbergstr. 4, 85747 Garching, Germany}
\altaffiliation{Fritz-Haber-Institut der Max-Planck-Gesellschaft, Faradayweg 4-6, 14195 Berlin, Germany}
\author{Harald Oberhofer}
\affiliation{Chair for Theoretical Chemistry and Catalysis Research Center, Technische Universit\"at M\"unchen, Lichtenbergstr. 4, 85747 Garching, Germany}

\title{Piece-wise multipole-expansion implicit solvation for arbitrarily shaped molecular solutes}

\begin{document}

\maketitle

\begin{abstract}
The multipole-expansion (MPE) model is an implicit solvation model used to efficiently incorporate solvent effects in quantum chemistry. Even within the recent direct approach, the multipole basis used in MPE to express the dielectric response still solves the electrostatic problem inefficiently or not at all for solutes larger than $\approx 10$ non-hydrogen atoms. In existing MPE parameterizations, the resulting systematic underestimation of the electrostatic solute-solvent interaction is presently compensated for by a systematic overestimation of non-electrostatic attractive interactions. Even though the MPE model can thus reproduce experimental free energies of solvation of small molecules remarkably well, the inherent error cancellation makes it hard to assign physical meaning to the individual free energy terms in the model, raising concerns about transferability. Here, we resolve this issue by solving the electrostatic problem piece-wise in 3D regions centered around all non-hydrogen nuclei of the solute, ensuring reliable convergence of the multipole series. 
The resulting method\note{removed} thus allows for a much improved reproduction of the dielectric response of a medium to a solute. Employing a reduced non-electrostatic model with a single free parameter, in addition to the density isovalue defining the solvation cavity, \change{our method} 
yields free energies of solvation of neutral, anionic and cationic solutes in water in good agreement with experiment.
\end{abstract}

\section{Introduction \label{sec:intro}}

With many---if not most---chemical processes of interest in biochemistry\cite{waterProtein}, electrochemistry\cite{reviewWaterSplitting,aqueousBatteries}, catalysis\cite{reviewSolventEffectsCat} and related fields\cite{reviewDrugDelivery} taking place in solution or at liquid interfaces with other phases, it is not surprising that the modelling of liquid environments is a major topic in computational chemistry.\cite{reviewSolventEffectsCat,reviewAndreussiFisicaro,ChemRev:SolvationModels,schwarz2020electrochemical,ringe2021implicit} The fundamental issue is that, from a first-principles point of view, the separation of a liquid system into one or multiple `solutes' and a `solvent' is artificial. Electrons and nuclei of the solvent and the solute follow the same quantum mechanical (QM) and statistical laws, and the straightforward approach would be to treat both \change{at} an explicit, quantum mechanical level.\cite{reviewAndreussiFisicaro} In practice, however, treating both \change{at} the same level of theory quickly turns out as computationally intractable already for rather trivial systems. The numbers of electrons and nuclei in this complete picture are much larger than those of the solute alone, as usually a large number of solvent molecules would have to be included in the model system to obtain converged results. In addition, the solvent typically consists of small, highly mobile molecules. While some short-range ordering may occur, there is generally no long-range order and a thermodynamic sampling would be necessary to accurately account for all solvent degrees of freedom.\cite{reviewSolventEffectsCat,reviewAndreussiFisicaro}

This great effort to accurately account for solvent effects is in stark contrast to the empirically known fact that typically, if one is interested in the properties of a molecular system in solution (but not the properties of the solution under the influence of the solute), the solvent plays only a minor role compared to the atomic structure of the solute itself.\cite{reviewAndreussiFisicaro} It is therefore reasonable---and necessary---to simulate only the solute \change{at} a detailed, quantum mechanical level and treat the solvent merely as an environment for the solute, using a simpler model.\cite{ChemRev:SolvationModels} 

There are, of course, different practical approaches to this general idea.\cite{reviewSolventEffectsCat,schwarz2020electrochemical} One highly popular group of methods, which we focus on in this work, are the so-called implicit solvation or continuum embedding models.\cite{ChemRev:SolvationModels,reviewAndreussiFisicaro,ringe2021implicit} The general idea is to ignore any atomistic detail of the solvent altogether---although in some cases one or few individual solvent molecules may still need to be included explicitly.\cite{SM6,SMD} Instead, the time-averaged effect of the solvent on the solute is modelled using the former's macroscopic properties, most notably its relative dielectric permittivity, and/or empirical models fitted to experimental reference data, typically free energies of solvation in the limit of infinite dilution.\cite{MinSol}

At the heart of many implicit solvent models stands the interaction of the solute charge density with a dielectric medium. The difference between the individual methods lies mostly in the way this electrostatic problem is solved in practice. The here discussed multipole-expansion (MPE) methods expand the dielectric response of the solvent in a multipole series. Earlier versions\cite{Kirkwood1934,Onsager1936,RIVAIL1976,Rinaldi1983,DILLET1993,Rinaldi2004} can solve the electrostatic problem exactly only when
the charge density is completely localized within the cavity,\cite{Sinstein_thesis} which is the region around the solute assumed to be inaccessible to the solvent. This leads to the so-called outlying charge error when solute charge density and model dielectric overlap, as is often necessary to accurately describe the strong interaction with solvents such as water.\cite{Sinstein,nonel}

This rather severe limitation of the MPE method was recently lifted through a novel regularization of the electrostatic potential and the so-called direct approach by Sinstein \textit{et al.}\cite{Sinstein,Sinstein_thesis}. This generalization now allows the modelling of stronger electrostatic solute-solvent interactions, opening up the method for a more general use. Unfortunately, though, practical applications of both the original and
the direct approach of the MPE model so far suffered from convergence and accuracy issues for larger solutes. The reason for this is that in these models the expansion of the electrostatic potential inside the cavity relies on a single expansion center, typically in the geometric center of the solute. Thus solutes that deviate strongly from a roughly spherical shape tend to necessitate rather high multipole expansion orders.\cite{Sinstein} Indeed in this work, we show that for some solutes the multipole series might fail to converge altogether. As a remedy, we here present a new approach, based on the partition of space into small subcavities, \change{similar in philosophy to the domain decomposition conductor-like screening model (ddCOSMO)\cite{ddcosmo1,ddcosmo2,ddcosmo3}. This adaptation now allows for} a true multi-center expansion of the electrostatic potential. We show that this new variant of the MPE model lifts size restrictions of the solute and allows for a much more stable convergence with expansion order.

This work is organized as follows: After briefly recapitulating the current state of the art of the MPE method in \cref{sec:Theory}, we thoroughly test the model on a popular benchmark set\cite{MinSol} described in \cref{sec:refdata}, in \cref{sec:benchmark_orig}. We develop the modified MPE subcavities method in \cref{sec:subcav} and demonstrate its efficacy in \cref{sec:results}. There we further present a new parameterization of the model for water as a solvent. Lastly, we discuss how the success of the original method\cite{Sinstein} in describing solution in water is based on a size-dependent systematic error cancellation. Our modified method is largely free of such phenomena, improving transferability.

\section{\label{sec:Theory}Theoretical and technical background}

\subsection{\label{sec:Scope}Scope of the model}

We consider a cluster-like (i.e.~not periodic in any real-space direction) molecular system (henceforth referred to as `solute') surrounded by a liquid solvent. One of the main characteristics of the solvent in any implicit solvation model is its dielectric permittivity. Therefore, the terms `solvent' and `dielectric (medium)' are used interchangeably in the present work. In contrast, the term `(in) vacuum' will refer to the absence of a solvent. The solute is treated \change{at} a density-functional theory (DFT) level of theory\cite{Sinstein,aims-paper,aims-Havu2009,aims-Yu2018,aims-Ren2012,aims-Ihrig2015}, while the molecules of the solvent are not treated explicitly. Instead, their time-averaged dielectric response is approximated in a classical electrostatic model.\cite{Sinstein,ChemRev:SolvationModels,reviewAndreussiFisicaro,ringe2021implicit} \change{Space is divided into a finite number of regions $X_i$, where the index $i$ can take arbitrary values, depending on the specific problem definition. Each $X_i$ is} assigned to either the dielectric medium or vacuum, with the latter describing the inside of the so-called solvation cavity, i.e. the region around the solute which is assumed to be inaccessible by solvent molecules. Any dielectric effects within this solvation cavity are explicitly accounted for at the DFT level of theory. Within the region belonging to the implicit dielectric, the dielectric response is instead assumed to be local, linear, homogeneous and isotropic. It can thus be described by a scalar relative permittivity $\varepsilon_i$, which is constant within a particular dielectric. 
\begin{align}\label{eq:epsilon_i_def}
	\varepsilon_{i} = \varepsilon(\mathbf{r} \in X_{i}) = \rm const. \quad .
\end{align}
There are, however, transitions between the different dielectric regions. There, the permittivity may show a complicated spatial behavior,\cite{bonthuis2013beyond} which implicit models tend to replace with simple smeared out step-functions for the sake of efficiency.\cite{nonel,fisicaro2017soft,solvent_aware_interface} In this spirit, the MPE method, together with other popular approaches such as COSMO\cite{klamt1993cosmo}, SMx\cite{SMD,SM6}, or the original PCM model\cite{ChemRev:SolvationModels},
defines a sharp transition on the 2D boundaries $\change{B}_{ij}$ between 3D regions $X_i$ and $X_j$.
In \cref{sec:subcav} below we will generalize this concept to an arbitrary number of regions in space $X_i$. In the following, we therefore keep the discussion at the level of general regions $X_i$ which obey \cref{eq:epsilon_i_def}. The exact definition of their shape will be given in the next section.

Finally, we restrict ourselves to the time-independent and non-periodic case in the absence of external fields other than that of the nuclei for the time being. Furthermore, the solvent contains no ions (the solute may, however, be ionic). This means that ionic liquids as well as saline solutions\cite{Ringe} lie beyond the scope of the present model.

\subsection{\label{sec:Electrostatics}Electrostatic problem}

The electrostatic problem is solved identically to previous work, using the ansatz\cite{Sinstein}
\begin{equation}\label{eq:ansatz}
	\Phi(\mathbf{r})=\varepsilon^{-1}(\mathbf{r})\Phi_\text{H}(\mathbf{r})+\Phi_\text{MPE}(\mathbf{r}) \quad ,
\end{equation}
where the electrostatic potential $\Phi_\text{MPE}$ is due to the solvent response. $\Phi_\text{H}$ is the classical Coulomb---or, in a DFT\cite{Sinstein,aims-paper,aims-Havu2009,aims-Yu2018,aims-Ren2012,aims-Ihrig2015} context, Hartree---potential\cite{electrostatics-textbook} of the solute charge density $\rho$
\begin{equation}\label{eq:Phi_H_integral}
	\Phi_\text{H}(\mathbf{r})=\int\limits_{\mathbb{R}^3} d^3\mathbf{r}^\prime\,  \frac{\rho(\mathbf{r}^\prime)}{\|\mathbf{r}-\mathbf{r}^\prime\|} \quad .
\end{equation}
The latter obeys Poisson's equation in vacuum,
\begin{equation}
	\nabla^2\Phi_\text{H}(\mathbf{r}) = -4\pi\rho(\mathbf{r}) \quad .
\end{equation}
The net charge density $\rho$ includes both electrons and nuclei of the solute,
\begin{equation}
	\rho(\mathbf{r}) = - \rho_\text{el}(\mathbf{r}) + \sum_{N=1}^{N_\text{max}} Q_N \delta(\mathbf{r}-\mathbf{R}_N) \quad .
\end{equation}
Here, $\rho_\text{el}$ is the solute electron density, $\delta$ is the delta function, and $\mathbf{R}_N$ and $Q_N$ are respectively the nuclear coordinates and charges of the $N_{\rm max}$ nuclei constituting the explicitly treated solute.\cite{Sinstein}

As mentioned above, implicit solvation models generally partition space into a dielectric medium $X_{\rm Q}$ and a so-called `cavity' $X_{\rm R}$ in which the solute resides.\cite{Sinstein,ChemRev:SolvationModels,reviewAndreussiFisicaro,ringe2021implicit} For these regions, the relative dielectric permittivity is defined by
\begin{subequations}
\begin{align}
    \varepsilon(\mathbf{r}\in X_{\rm R}) &= 1 \label{eq:epsilon_in_R} \\
    \varepsilon(\mathbf{r}\in X_{\rm Q}) &= \varepsilon_{\rm bulk} \quad , \label{eq:epsilon_in_Q}
\end{align}
\end{subequations}
where $\varepsilon_{\rm bulk}$ is the experimentally determined macroscopic permittivity of the solvent. The spatial extents of these regions are defined by an isovalue $\rho_\text{iso}$ of the solute electronic density through
\begin{subequations}
\begin{align}
    X_\text{R} &= \{ \mathbf{r}: \widetilde{\rho}_\text{el}(\mathbf{r}) > \rho_\text{iso} \} \label{eq:def_cavity_orig} \\
    X_\text{Q} &= \{ \mathbf{r}: \widetilde{\rho}_\text{el}(\mathbf{r}) < \rho_\text{iso} \} \label{eq:def_dielectric_orig}
\end{align}
\end{subequations}
\change{An electronic density isosurface is per definition smooth, continuous and follows the shape of the solute. It is thus a convenient way to define solvation cavities and is commonly used in various implicit solvation models, implemented in different electronic structure packages.\cite{nonel, Ringe, Sinstein, vaspsol}} The isovalue $\rho_\text{iso}$ thereby takes the form of a free parameter that needs to be determined e.g.~by fitting to a suitable training set, cf.~\cref{sec:parameterization}. \change{However, other definitions, e.g. based on atom-centered spheres\cite{SMD,SSCS,ddcosmo1}, are also commonly used. The method described in \cref{sec:subcav} makes no assumptions about the cavity shape other than that it roughly follows the shape of the solute and that it is sharp in the sense of \cref{eq:epsilon_in_R,eq:epsilon_in_Q}. It can thus, in principle, be used together with any such cavity definition. An electronic density isosurface is merely the choice which we use to demonstrate the method in the present work.} 

\note{skip added}

Being an effective model, though, the exact choice of which electron density $\widetilde{\rho}_{\rm el}$ to use in the above definition is ambiguous. \change{Different} possibilities have been proposed\cite{Sinstein,Sinstein_thesis}\change{.\note{shortened} In the present work we use the superposition of free atom electron densities $\rho_{\rm free}$, i.e.~the electron densities of the solute's individual atoms in vacuum\cite{aims-paper}. We refer the reader to the cited publications for the other options. It has been shown before\cite{Sinstein,Sinstein_thesis} that those choices lead to a systematic error in the free energy of solvation for anionic solutes. The superposition of $\rho_{\rm free}$ does not suffer from this issue and is able to describe neutral, cationic and anionic solvents all with the same set of parameters\cite{Sinstein_thesis}. It should be noted here that in practice, when using a hybrid exchange correlation functional, the DFT software \aims which we use throughout the present work calculates $\rho_{\rm free}$ from its generalized gradient approximation (GGA) part, so e.g. the PBE\cite{PBE} and HSE06\cite{hse03,hse06} functionals will in practice yield the exact same cavity.}

Note that the non self-consistent electron density $\widetilde{\rho}_\text{el}$ is used solely for the cavity definition. The electrostatic potentials $\Phi_{\rm H}$ and $\Phi_{\rm MPE}$, which depend on the former, are, of course, calculated self-consistently from the instantaneous electron density $\rho_{\rm el}$ of the solute at each SCF step.

Next, we need to choose a basis to represent the solvent response to the electrostatic potential given in \cref{eq:ansatz}.\note{shortened} \change{It has been shown that in a region of constant $\varepsilon(\mathbf{r})$, $\Phi_{\rm MPE}$ is a harmonic function, lending itself to a series expansion in regular solid harmonic functions $\mathcal{R}_m^l$ or their irregular counterparts $\mathcal{I}_m^l$.}\cite{Kirkwood1934,Onsager1936,RIVAIL1976,Rinaldi1983,DILLET1993,Rinaldi2004,Sinstein}
\begin{subequations}
\begin{align}
    \mathcal{R}_m^l(\mathbf{r}) &=  r^l \, Y_m^l(\theta,\varphi) \label{eq:solharm_R} \\
    \mathcal{I}_m^l(\mathbf{r}) &= r^{-(l+1)} \, Y_m^l(\theta,\varphi)  \\
    l,m \in \mathbb{Z}&, \quad l \geq 0, \quad -l \leq m \leq l \nonumber
\end{align}
\label{eq:solidharm}
\end{subequations}
Here, $Y_m^l$ are the real valued spherical harmonics (cf.~ref.~\citenum{aims-paper}). We\note{removed} thus use a piece-wise definition of $\Phi_\text{MPE}$, treating each region $X_i$ separately.\cite{Sinstein} We define
\begin{equation}
    \Phi_i=\Phi_\text{MPE}(\mathbf{r}\in X_i) \quad .
\end{equation}
If $X_i$ is bounded, i.e.~if
\begin{equation}\label{eq:bounded}
	\exists r\in\mathbb{R}^+ :\, r > \|\mathbf{r}\| \, \forall \mathbf{r} \in X_i \quad ,
\end{equation}
as is the case inside the solvation cavity $X_{\rm R}$, then\note{removed} $\Phi_i$ can be expanded in a set $\{\mathcal{R}_m^l\}_K$ of ${\mathcal{R}_m^l(\mathbf{r}-\mathbf{r}_K)}$ around a center $\mathbf{r}_K$\cite{Kirkwood1934,Onsager1936,RIVAIL1976,Rinaldi1983,DILLET1993,Rinaldi2004,Sinstein}
\begin{equation}\label{eq:expansion_R_general}
	\Phi_{i}(\mathbf{r}) = \sum_{l=0}^{l^\text{R}_{\text{max}}} \sum_{m=-l}^l R_K^{(l,m)} \mathcal{R}_m^l(\mathbf{r}-\mathbf{r}_K) \quad ,
\end{equation}
with expansion coefficients $R_K^{(l,m)}$. If $X_i$ is unbounded, i.e.~if \cref{eq:bounded} is not satisfied, as is the case in $X_{\rm Q}$, then\note{removed} $\Phi_i$ can be expanded in a union of sets $\{\mathcal{I}_m^l\}_J$ of $\mathcal{I}_m^l(\mathbf{r}-\mathbf{r}_J)$ with multiple centers $\mathbf{r}_J$\cite{Kirkwood1934,Onsager1936,RIVAIL1976,Rinaldi1983,DILLET1993,Rinaldi2004,Sinstein}
\begin{equation}\label{eq:expansion_Q_general}
	\Phi_i(\mathbf{r}) = \sum_{J=1}^{J_\text{max}} \sum_{l=0}^{l^\text{Q}_{\text{max}}} \sum_{m=-l}^l Q_J^{(l,m)} \mathcal{I}_m^l(\mathbf{r}-\mathbf{r}_J) \quad ,
\end{equation}
and with expansion coefficients $Q_J^{(l,m)}$. The only \change{singularities} are \change{at} $\mathbf{r}=\mathbf{r}_J$. By choosing $\mathbf{r}_J \notin X_i\, \forall J$, \change{$\Phi_i$ is continuous in } $X_i$ \change{by construction}.\cite{Kirkwood1934,Onsager1936,RIVAIL1976,Rinaldi1983,DILLET1993,Rinaldi2004,Sinstein}

$\Phi_{\rm MPE}$ can then be uniquely defined by imposing continuity of the total electrostatic potential and of the electric flux density at $\change{B}_{ij}$.\cite{ChemRev:SolvationModels} Together with the regularization in \cref{eq:ansatz}, this leads to the following boundary conditions\cite{Sinstein}
\begin{subequations}
\label{eq:continuity_boundary_general}
\begin{align}
	\Phi_{i}(\mathbf{r}) - \Phi_{j}(\mathbf{r}) &
	= \left( \varepsilon_{j}^{-1} - \varepsilon_i^{-1} \right) \Phi_\text{H}(\mathbf{r}) \label{eq:continuity_boundary_general_pot} \\
	\mathbf{n}_\mathbf{r} \nabla \left( \varepsilon_{i}\Phi_{i}(\mathbf{r}) - \varepsilon_{j}\Phi_{j}(\mathbf{r}) \right) &
	= 0 \label{eq:continuity_boundary_general_field} \\
	\forall \mathbf{r} \in \change{B}_{ij}, &\, \forall i\neq j \quad ,\nonumber
\end{align}
\end{subequations}
where $\mathbf{n}_\mathbf{r}$ is the surface normal of $\change{B}_{ij}$ at $\mathbf{r}$.

\change{We insert \cref{eq:expansion_R_general} and \cref{eq:expansion_Q_general} into \cref{eq:continuity_boundary_general}, for bounded and unbounded $X_{i/j}$, respectively. Evaluating it} at sets of points situated at the interfaces $\change{B}_{ij}$ then essentially turns the problem of determining the dielectric response 
into an algebraic problem of solving an overdetermined system of linear equations (SLE) for the expansion coefficients of the potential\note{removed}\cite{Sinstein,Rinaldi1983}.
\change{
\begin{equation}\label{eq:SLE}
    \mathbf{Ax}=\mathbf{b}
\end{equation}
Here, $\mathbf{A}$ contains the basis functions $\mathcal{R}_m^l$ and $\mathcal{I}_m^l$ for \cref{eq:continuity_boundary_general_pot} and their scaled derivates for \cref{eq:continuity_boundary_general_field}, evaluated at said discrete set of points on $B_{ij}$. $\mathbf{b}$ contains the scaled Hartree potential for \cref{eq:continuity_boundary_general_pot} and zero for \cref{eq:continuity_boundary_general_field}. The solution vector $\mathbf{x}$ contains the expansion coefficients $R_K^{(l,m)}$ and $I_J^{(l,m)}$. The discretization algorithm for $B_{ij}$ aims to achieve a certain degree of determination $d_{\rm det}$, i.e. the ratio between rows and columns in $\mathbf{A}$. \Cref{eq:SLE}} can finally be solved using a number of algorithms which all more or less reduce to a linear least-squares fit\note{removed}. \change{The technical details of this numeric solution have been described in ref.~\citenum{Sinstein}.} As a measure for the quality of the solution to the linear algebra problem we use the adjusted coefficient of determination $\Bar{R}^2$\note{removed},\change{\cite{statistics,statistics_paper,wherry1931}} which essentially measures how much of the variability of the dependent variable of an equation is due to a linear relationship with the \change{in}dependent variable. \change{The calculation of $\Bar{R}^2$ is described in the supporting information (SI) of the present work.}

With $\Phi_{\rm MPE}$ defined in this manner, it is possible to calculate the free energy $G^{\rm elstat}(\{\mathbf{R}_N\})$ of a molecule electrostatically embedded into the solvent by replacing the Hartree potential with \cref{eq:ansatz} in a DFT\cite{Sinstein,aims-paper,aims-Havu2009,aims-Yu2018,aims-Ren2012,aims-Ihrig2015} calculation. Using the total energy of the solute in vacuum $E_{\rm vac}(\{\mathbf{R}_N\})$, this then defines the electrostatic contribution to the solvation free energy of the solute, $\Delta G_{\rm solv}^{\rm elstat}(\{\mathbf{R}_N\}) =  G^\text{elstat} (\{\mathbf{R}_N\}) - E_\text{vac}(\{\mathbf{R}_N\})$.
Note that all these quantities refer to the respective electronic ground state and we have here explicitly written the dependence on the nuclear positions $\{\mathbf{R}_N\}$. For clarity of notation, this dependence will henceforth be dropped.

\subsection{\label{sec:nonel_theo}Non-electrostatic free energy contributions}

There are, of course, further contributions to the free energy of a solute in solvent, beyond the hitherto discussed electrostatics. Specifically, solute and solvent also interact via Pauli repulsion and dispersion forces, and there are entropic and even grand-canonical contributions to consider due to the displacement of solvent molecules (cavity formation). We refer to recent reviews for a detailed overview over these, so-called non-electrostatic contributions generally considered in implicit solvation models.\cite{ChemRev:SolvationModels,schwarz2020electrochemical,ringe2021implicit} Here, we only point out that these different contributions are often lumped together into effective expressions with a minimum number of free parameters. The solvation free energy is correspondingly written as arising from two contributions, 
\begin{equation}\label{eq:model_dG_solv}
    \Delta G_\text{solv} = 
    \Delta G_{\rm solv}^{\rm elstat} + 
    \Delta G_{\rm solv}^{\rm  non-elstat} \quad ,
\end{equation}
with $G_{\rm solv}^{\rm non-elstat}$ containing all non-electrostatic contributions. One frequent functional expression for the latter is a simple linear function of surface area $A$ and volume $V$ of the solvation cavity\cite{nonel}
\begin{equation}\label{eq:nonel_Sinstein}
    \Delta G_\text{solv}^\text{non-elstat} = \alpha A + \beta V \quad ,
\end{equation}
with $\alpha$ and $\beta$ empirical parameters that are fitted to a set of experimental free energies of solvation. While this functional form was initially established for the self-consistent continuum solvation (SCCS) model,\cite{nonel} it turns out that such a treatment of non-electrostatics can also be used with other electrostatic models such as MPE.\cite{Sinstein} 

Both $A$ and $V$ are a measure of molecular size and are therefore typically correlated. Considering the effective character of the functional form employed for the non-electrostatic free energy contribution, it is correspondingly possible to also use only one of its two terms, typically the surface area one, $\alpha A$, at generally insignificant increases in the mean error of computed solvation free energies with respect to experimental references.\cite{nonel,Sinstein,Sinstein_thesis} For our present purpose of assessing and removing error cancellation between the electrostatic and non-electrostatic contributions to $\Delta G_{\rm solv}$, having less flexibility in a non-electrostatic term with only one fit parameter is actually even beneficial. In this work, we correspondingly use as simple non-electrostatic model,
\begin{equation}\label{eq:nonel_only_A}
    \Delta G_\text{solv}^\text{non-elstat} = \alpha A \quad .
\end{equation}
Following earlier work,\cite{Sinstein} we also treat $\Delta G_\text{solv}^\text{non-elstat}$ as a post-SCF correction instead of including it self-consistently in the Kohn-Sham\cite{KS-paper} (KS) operator.

It is worth noting that in fitting the parameter $\alpha$ to experimental reference data\cite{MinSol}, see below, the free energy contributions due to nuclear degrees of freedom are to some degree accounted for in $\Delta G_\text{solv}^\text{non-elstat}$, at least in a statistical manner.\cite{Sinstein_thesis} For unpolar organic solvents, the problem may also emerge that the optimized $\alpha$ becomes negative. With $\Delta G_\text{solv}^\text{elstat}$ negative by construction and then $\Delta G_\text{solv}^\text{non-elstat}$ also negative, the solvation free energy would necessarily result as exothermic. The case of a solute being repelled by the solvent thus lies outside the scope of such a model. Notwithstanding, in the present work we focus on solvation in water, where $\alpha$ is typically positive\cite{nonel,Sinstein} and this problem does not play a role.

\subsection{\label{sec:tec}Technical details}

The computational setup used to calculate $\Delta G_{\rm solv}^{\rm elstat}$ is identical to the one of preceding work and we refer to the corresponding publication for all technical details\cite{Sinstein}. Unless mentioned otherwise, all DFT calculations are performed with \aims \change{(version \texttt{210415})} into which the present implicit solvent model was implemented\note{moved}. \change{The} \emph{tight} default\note{removed} electronic basis sets and\note{removed} integration grid\change{s} included in the \aims package are employed throughout. We use the Perdew, Burke, Ernzerhof (PBE) exchange-correlation (xc) functional\cite{PBE} with collinear spin and an atom-wise scalar zeroth order regular approximation (ZORA)\cite{aims-paper} for relativistic effects of the core electrons. Kohn-Sham levels are occupied through a Gaussian broadening scheme\cite{Gaussian_broadening} with a rather narrow width of $0.01$eV to account for the molecular nature of our solutes. \change{\Cref{eq:SLE}} was solved in a direct solver by QR factorization of the left-hand side matrix and subsequent singular value decomposition of the R matrix,\cite{ScaLAPACK-doc,LAPACK-doc} as described before\note{removed}\cite{Sinstein}.

\section{Benchmark systems and experimental reference data \label{sec:refdata}}

All molecular geometries and experimental reference values for $\Delta G_{\rm solv}$ used in the present work were taken from the Minnesota Solvation Database, version 2012.\cite{MinSol}
We use all 790 small molecules included in the database to investigate the convergence of $\Delta G^{\rm elstat}_{\rm solv}$ with numeric parameters in the respective models, especially with the expansion order. \change{We furthermore use this set to test the computational performance of our method.} In order to keep these convergence studies as unbiased and to make the resulting parameters as generally applicable as possible, we randomize the two physical model parameters $\varepsilon$ and $\rho_{\rm iso}$. For each of the 790 test solutes, two random numbers $\mathfrak{R}_{\varepsilon/\rho}$ are drawn from a uniform distribution between $0$ and $1$.
The parameters are then calculated according to
\begin{subequations}
\begin{align}
    \varepsilon &= \frac{2}{\mathfrak{R}_\varepsilon} \\
    \rho_{\rm iso} &= 10^{-(1+3\cdot\mathfrak{R}_\rho)} \,\mathrm{e\,\si{\angstrom}}^{-3} \quad ,
\end{align}
\end{subequations}
yielding $\varepsilon>2$ with the distribution becoming increasingly sparse for higher values. A molecular geometry together with its $\varepsilon$ and $\rho_{\rm iso}$ will henceforth be referred to as a system.

While any $\varepsilon\geq 1$ is physically possible, this distribution is chosen to roughly mimic the range of permittivities of real-life solvents, which are rarely ever much smaller than $2$.\cite{MinSol}
The isodensities are uniformly distributed on a logarithmic scale, ranging from $10^{-1}$ to $10^{-4}\,\mathrm{e\,\si{\angstrom}}^{-3}$ which is roughly the range in which we expect to find optimized isodensities for different solvents based on previous experience with the original direct version of the MPE model\cite{Sinstein,Sinstein_thesis}. Note that as the cavity tends to become smoother for lower isodensities, it is reasonable to assume that any set of converged parameters within the sampled isodensity range can also be applied to lower isodensities. In fact, we will see in the following section, specifically in \cref{tab:mols_unsolvable}, that systems for which the electrostatic problem is hard to solve typically have high $\rho_{\rm iso}$. To ensure comparability between calculations, the randomly chosen permittivities and isodensities differ only between solutes, but for each solute the same values are used for all convergence studies.

\change{DFT calculations with implicit solvation} may occasionally fail for various reasons. In our case, this happened mostly due to issues with the cavity surface discretization algorithm (cf. SI as well as ref.~\citenum{Sinstein}). Addressing these issues is beyond the scope of the present work. Due to such failures occurring only rarely, we simply ignore problematic systems for the time being. Unless mentioned otherwise, within each convergence study\change{, performance test} or parameterization, solutes for which at least one calculation failed for any reason are discarded from further analysis.

For the parameterizations in \cref{sec:parameterization} we used all solutes for which $\Delta G_{\rm solv}$ in water is available in the Minnesota Solvation Database. For some ionic solutes, a version clustered by one explicit water molecule and an unclustered version is available\cite{SM6,SMD}. For these solutes, only the clustered versions were used. This set of training data is slightly larger than in previous work where only some subsets of the database were used.\cite{Sinstein}

\section{Convergence of the original MPE model \label{sec:benchmark_orig}}

The original version of the direct MPE model\cite{Sinstein} partitioned space only into the cavity and the dielectric medium, as defined in \cref{eq:def_cavity_orig,eq:def_dielectric_orig}. The basis set for the dielectric response inside the cavity $\Phi_{\rm R}$ were the regular solid harmonics $\{\mathcal{R}_m^l\}$ centered around the geometric center of the molecule. Within the perspective of the present generalization, we will henceforth call this model MPE-1c, with 1c indicating the single cavity $X_{\rm R}$. Correspondingly, we will also denote electrostatic solvation free energies as $\Delta G_{\rm solv}^{\rm elstat}(n{\rm c})$ to indicate that they have been computed with an $n$ cavity model. In this section, we explore the capabilities and limitations of MPE-1c, making no conceptual modifications with respect to the original publication.

\subsection{Insufficiency of the solid harmonic basis \label{sec:original_insuff}}

To highlight the initially mentioned multipole expansion issues of the MPE-1c model, we analyze the convergence behavior of $\Delta G_{\rm solv}^{\rm elstat}(l_{\rm max}^{\rm R}; 1{\rm c})$ with increasing expansion order $l_{\rm max}^{\rm R}$ of $\Phi_{\rm R}$ using the test systems described in \cref{sec:refdata}. Specifically, we test \change{$l_{\rm max}^{\rm R}=2,4,6,8,12,16$ and $20$}. For $\Phi_{\rm Q}$, we use $l_{\rm max}^{\rm Q}=6$.\cite{Sinstein} As a reference, we use the $\Delta G^{\rm elstat}_{\rm solv}(l_{\rm max}^{\rm R}=20; n{\rm c})$ obtained for the same systems with our new method MPE-$n$c at $l_{\rm max}^{\rm R}=20$ and $l_{\rm max}^{\rm Q}=8$. Further explaining this new method in \cref{sec:subcav}, we show in \cref{sec:convergence} that these reference values themselves are converged up to on average $\lesssim 1\,{\rm meV}$ and thus constitute a firm reference. Additionally, we test the convergence of the adjusted coefficient of determination $\Bar{R}^2$\note{removed} of the solution to the discretized boundary conditions \cref{eq:continuity_boundary_general_pot,eq:continuity_boundary_general_field}.

\begin{figure}[ht!]
    \centering
    \includegraphics[width=\linewidth]{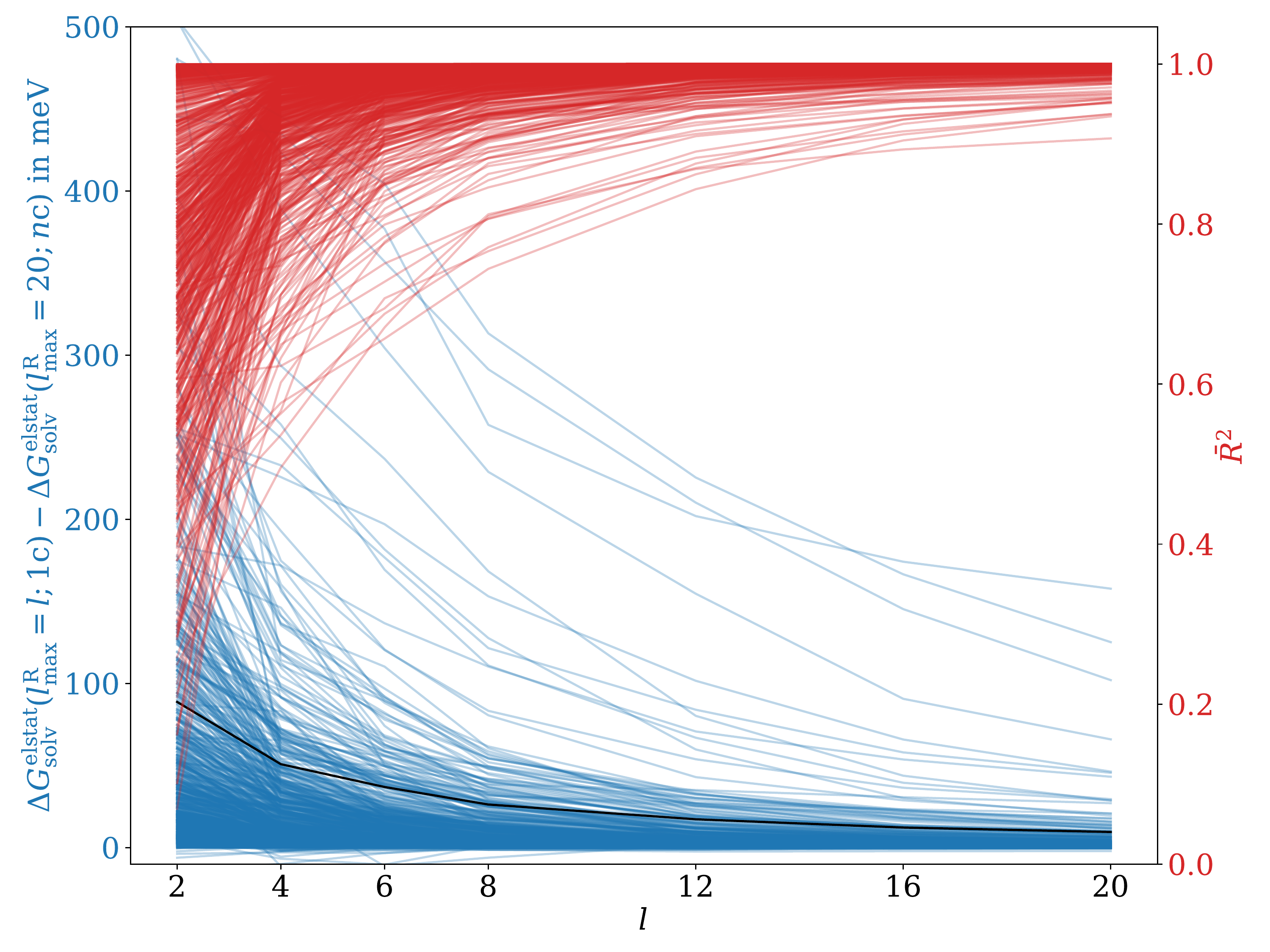}
    \caption{Convergence of the electrostatic contribution to the solvation free energy $\Delta G^{\rm elstat}_{\rm solv}(l_{\rm max}^{\rm R}; 1{\rm c})$ (blue) and adjusted coefficient of determination $\Bar{R}^2$ (red) of the discretized boundary conditions \cref{eq:continuity_boundary_general_pot,eq:continuity_boundary_general_field} with expansion order $l_{\rm max}^{\rm R}$ of the MPE potential inside the cavity $\Phi_{\rm R}$ in MPE-1c. Individual lines refer to individual systems contained in the test set as described in section 3. The root mean-square error over all systems is shown as a thick black line. As reference, we use $\Delta G_{\rm solv}^{\rm elstat}(l_{\rm max}^{\rm R}=20; n{\rm c})$ calculated with the MPE-$n$c model (see \cref{sec:subcav}), the same reference as also in \cref{fig:converge_lmax_rf}.\note{opacity of lines increased for better visibility}}
    \label{fig:converge_lmax_rf_nosubcav}
\end{figure}

The results are compiled in \cref{fig:converge_lmax_rf_nosubcav} and show clearly that MPE-1c systematically underestimates the electrostatic interaction between solute and solvent at the originally published ${l_{\rm max}^{\rm R}=8}$.\cite{Sinstein} Even at the significantly higher $l_{\rm max}^{\rm R}=20$, there are several test systems with errors in the order of $100\,{\rm meV}$. The root mean-square error (RMSE) at that expansion order, in contrast, is only $9.6\,{\rm meV}$. Below we will show that the larger errors are not outliers, but rather a systematic error occurring for larger solutes. The RMSE is only relatively low, because smaller molecules are over-represented in the test set.
\begin{figure}[ht]
    \centering
    \includegraphics[width=\linewidth]{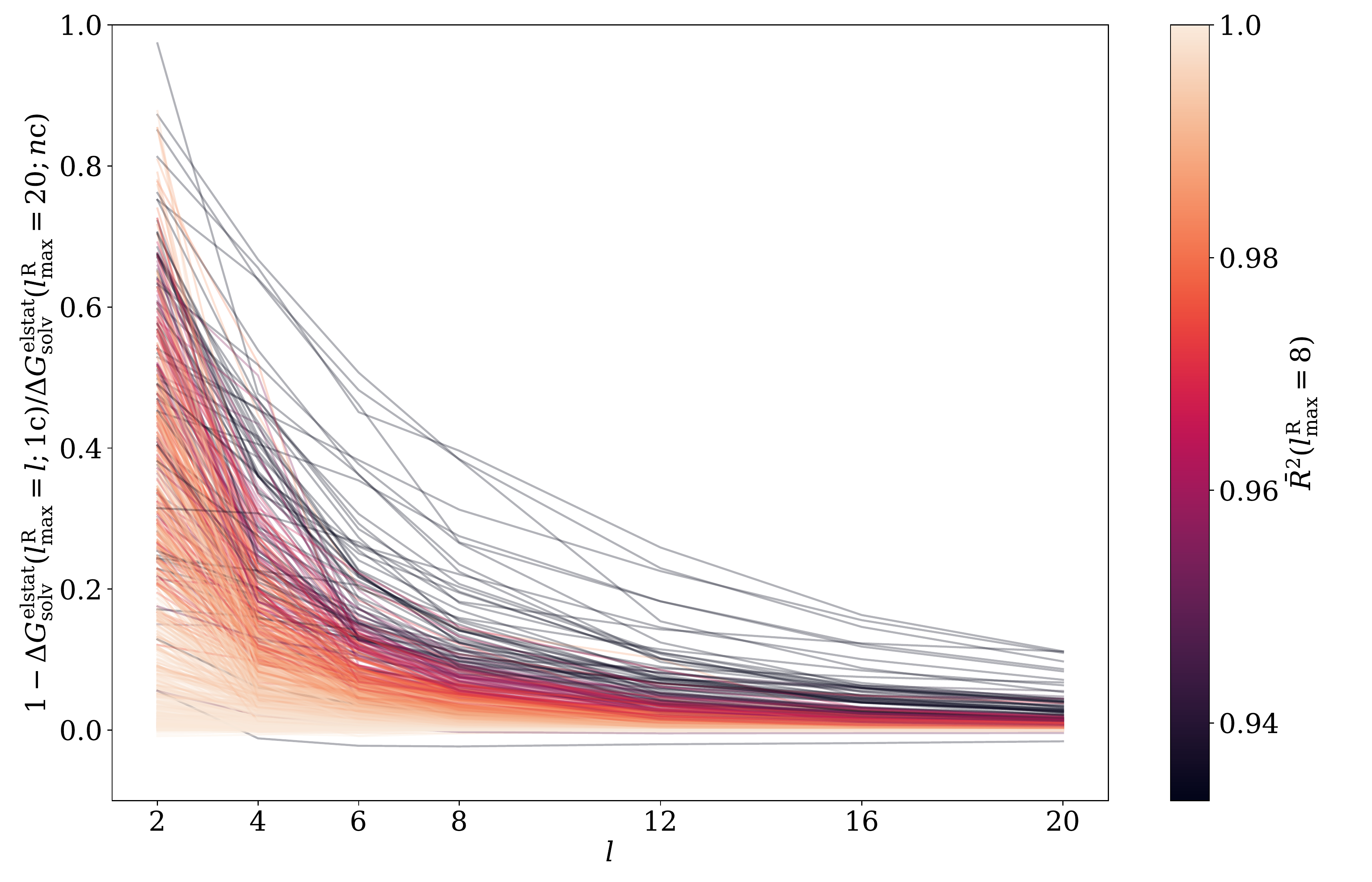}
    \caption{Relative error of the electrostatic contribution to the solvation free energy $\Delta G_{\rm solv}^{\rm elstat}({\rm 1c})$ in MPE-1c vs. the MPE-$n$c reference. Individual lines reflect the individual systems contained in the test set. The lines are colored according to the adjusted coefficient of determination $\Bar{R}^2$ of the discretized boundary conditions \cref{eq:continuity_boundary_general_pot,eq:continuity_boundary_general_field} at $l_{\rm max}^{\rm R}=8$, which is the originally published value for this parameter.\cite{Sinstein}\note{opacity of lines increased for better visibility}}
    \label{fig:converge_lmax_rf_nosubcav_rel}
\end{figure}
Furthermore, for some systems, even at $l_{\rm max}^{\rm R}=20$ the solid harmonic basis is insufficient to solve the electrostatic boundary conditions \cref{eq:continuity_boundary_general_pot,eq:continuity_boundary_general_field}, as indicated by $\Bar{R}^2<1$. It can straightforwardly be seen that the error in $\Delta G^{\rm elstat}_{\rm solv}$ is related to an insufficient solution of the electrostatic problem, by relating the relative error of $\Delta G_{\rm solv}^{\rm elstat}({\rm 1c})$ to $\Bar{R}^2$ as depicted in \cref{fig:converge_lmax_rf_nosubcav_rel}.

\begin{figure}[ht]
    \centering
    \includegraphics[width=\linewidth]{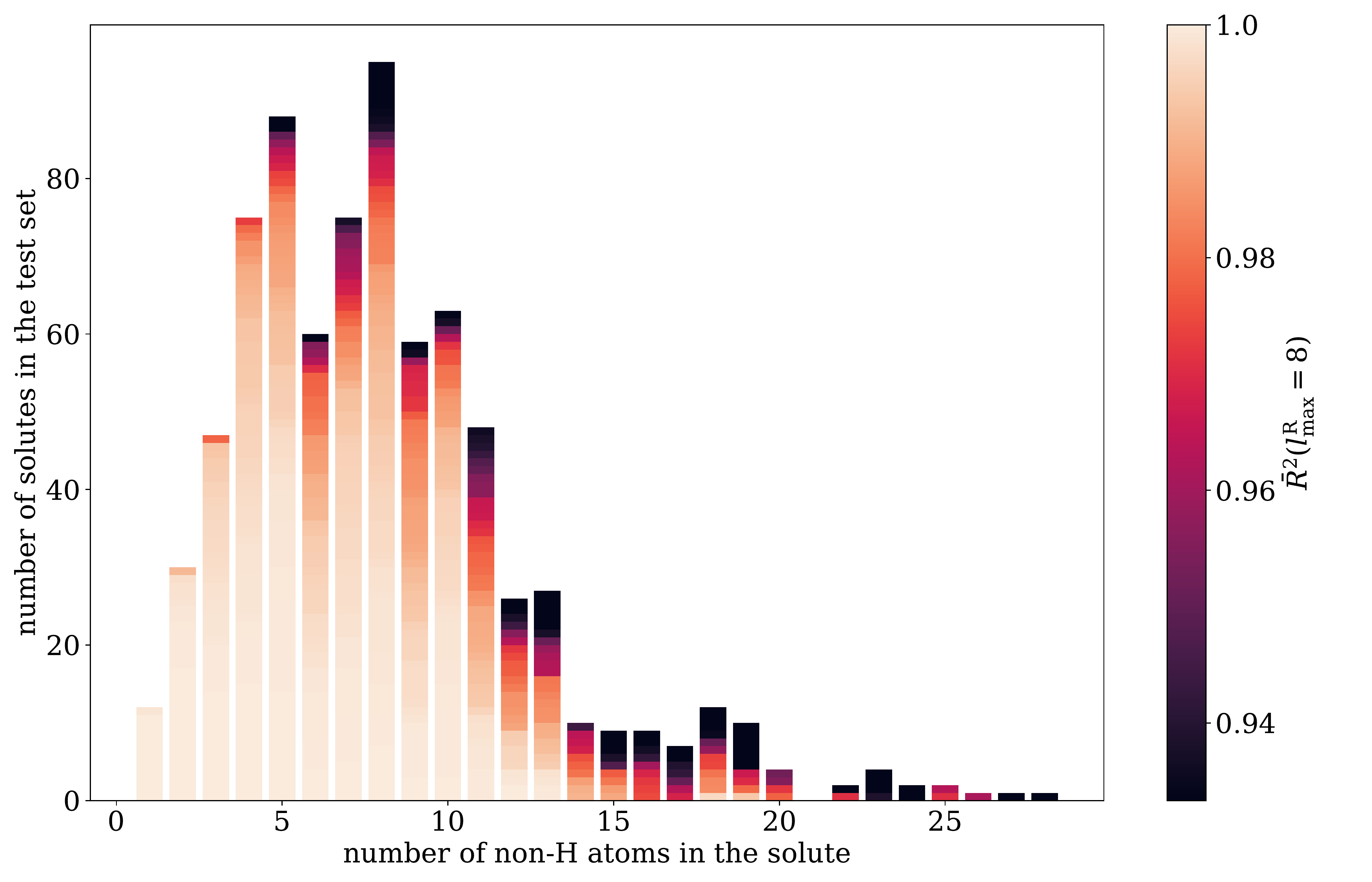}
    \caption{Histogram of the number of non-H atoms per solute in the test set, colored according to the adjusted coefficient of determination $\Bar{R}^2$ of the discretized boundary conditions \cref{eq:continuity_boundary_general_pot,eq:continuity_boundary_general_field} at $l_{\rm max}^{\rm R}=8$.}
    \label{fig:hist_n_nonH}
\end{figure}

Finally, \cref{fig:hist_n_nonH} shows that these errors in the electrostatic potential are related to the size of the molecules. The fraction of solutes with $\Bar{R}^2 \approx 1$ decreases with increasing number of \change{non-hydrogen (}non-H\change{)} atoms in the solute. Given the dominance of small molecules ($\leq 10$ non-H atoms), the relatively low RMSE in $\Delta G_{\rm solv}^{\rm elstat}$, apparent in \cref{fig:converge_lmax_rf_nosubcav}, is thus only representative for such smaller molecules, with increasingly larger errors to be expected for larger molecules.

Taking all of these findings into account, we conclude that MPE-1c with the originally published parameters\cite{Sinstein} suffers from a systematic underestimation of the electrostatic solute-solvent interaction. While negligibly small for small molecules, this error increases significantly for larger molecules.

\subsection{The wrong basis functions or too few? \label{sec:phyical_or_technical}}

An intuitive explanation why MPE-1c fails for larger molecules would be that arbitrary harmonic potentials $\nabla^2\Phi_\text{harm}(\mathbf{r}\in X)=0$ can be approximated to arbitrary accuracy in a finite series in $\{\mathcal{R}_m^l\}$, if $X$ is strictly convex. This follows from the translation theorem\cite{SolHarm1,SolHarm2} for arbitrary $\mathcal{R}_m^l$ being exact, truncated at the same order $l$, and converging everywhere in $\mathbb{R}^3$. For larger molecules, however, we expect the cavities $X_{\rm R}$ to deviate further from convex shapes. It is then not generally possible to know beforehand if the expansion will converge fast, or at all. It is worth noting here that except for the simplest systems, cavities are never perfectly convex in practice, yet most systems do actually converge.

Theoretical considerations about whether or not arbitrary harmonic potentials can be expanded in $\{\mathcal{R}_m^l\}$ in any given cavity $X_{\rm R}$ are of limited use. On one hand, even if we know that the infinite series converges, we do not generally know if it converges to desired accuracy at a practically applicable expansion order. On the other hand, even if harmonic functions exist for which the expansion does not converge, for $\Phi_{\rm R}$ in particular it may still converge.

\begin{figure}[ht]
    \centering
    \includegraphics[width=\linewidth]{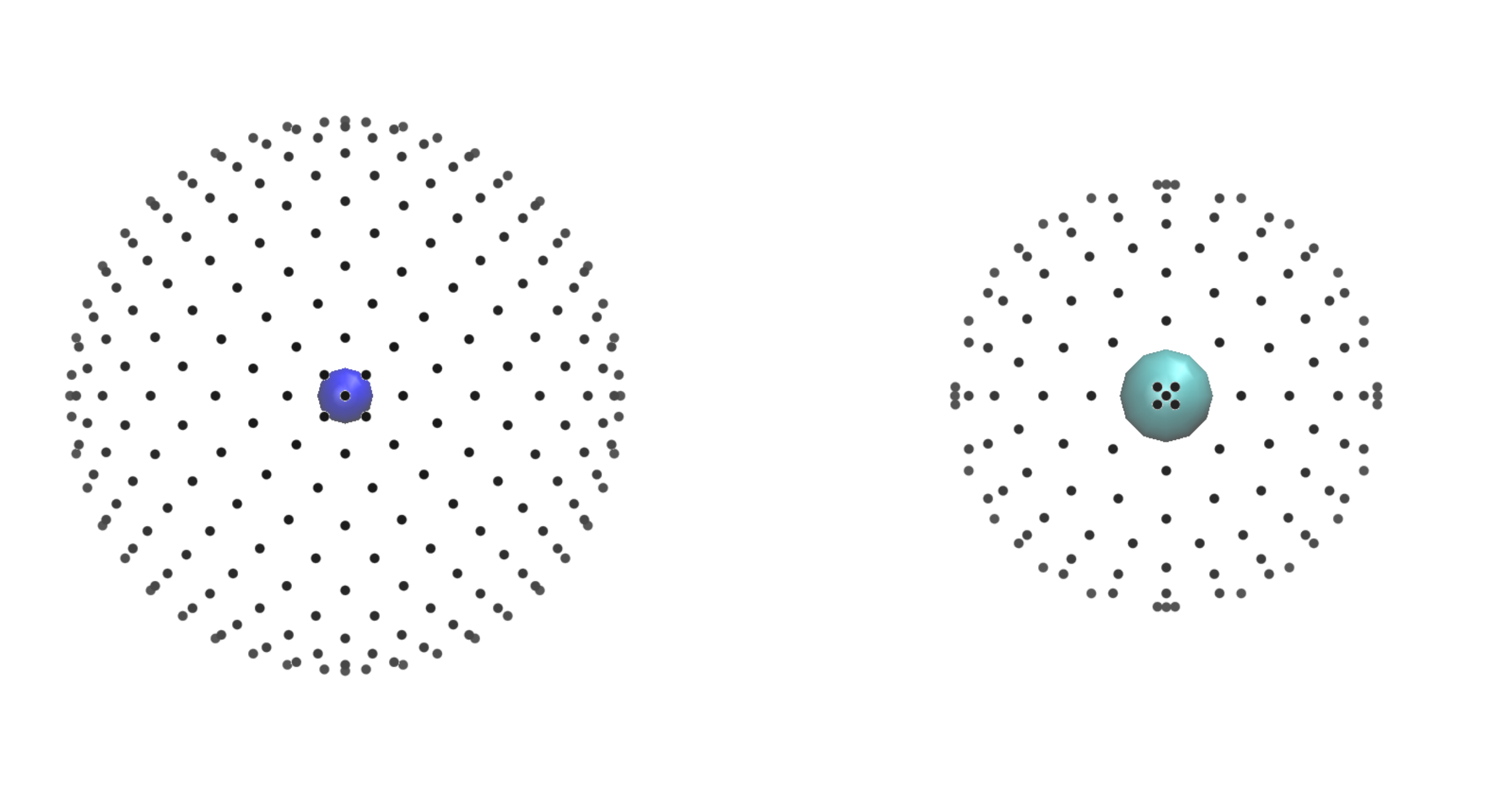}
    \caption{Na and Cl atom at $10\,\si{\angstrom}$ separation with discretized isodensity cavity surfaces at $\rho_{\rm iso}=0.01\,\frac{e}{\si{\angstrom}^3}$. Despite the two cavities being completely disjoint, one single expansion in regular solid harmonics around the center between the two is sufficient to solve the electrostatic problem. Note that by `one single expansion' we mean not only that the same basis is used in both cavities, but also the expansion coefficients are the same.}
    \label{fig:NaCl}
\end{figure}

The latter aspect can be illustrated by the following example: We consider one Na and one Cl atom at $10\,\si{\angstrom}$ separation, as shown in \cref{fig:NaCl}. We run a DFT calculation in solution with $\varepsilon=78.36$ and $\rho_{\rm iso}=0.01\,\frac{e}{\si{\angstrom}^3}$. Based on the above considerations, we would expect MPE-1c to fail for this system. The cavities of the two atoms are completely disjoint and there is no straightforward reason to assume that $\Phi_{\rm R}$ in both cavities could be described by the same multipole expansion with the same expansion coefficients. In practice, however, at $l_{\rm max}^{\rm R}=8$, we get a $\Bar{R}^2=0.9957$, which suggests that the electrostatic boundary conditions are almost exactly fulfilled. When we use separate multipole expansions for $\Phi_{\rm R}$ in both cavities (a specific case of the general method described in the following sections), we, of course, obtain a $\Bar{R}^2=1$, but the error in the electrostatic contribution to the solvation free energy made by the single expansion in comparison to the separate expansion is only 12\,meV. Using two different multipole expansions effectively doubles the number of basis functions for $\Phi_{\rm R}$, from 81 to 162 (although at each point $\mathbf{r}$ only 81 are used, they still amount to 162 degrees of freedom in \change{\cref{eq:SLE}}). Going back to the single expansion, we can use $l_{\rm max}^{\rm R}=12$ to reach 169 basis functions. With this, we then get $\Bar{R}^2=0.9982$ and the error with respect to the two-center expansion is reduced to a mere 1.8\,meV. By simply increasing the number of basis functions to approximately the same amount, we have thus achieved almost the same improvement as by using separate multipole expansions in both cavities.

\begin{figure}[ht]
    \centering
    \includegraphics[width=\linewidth]{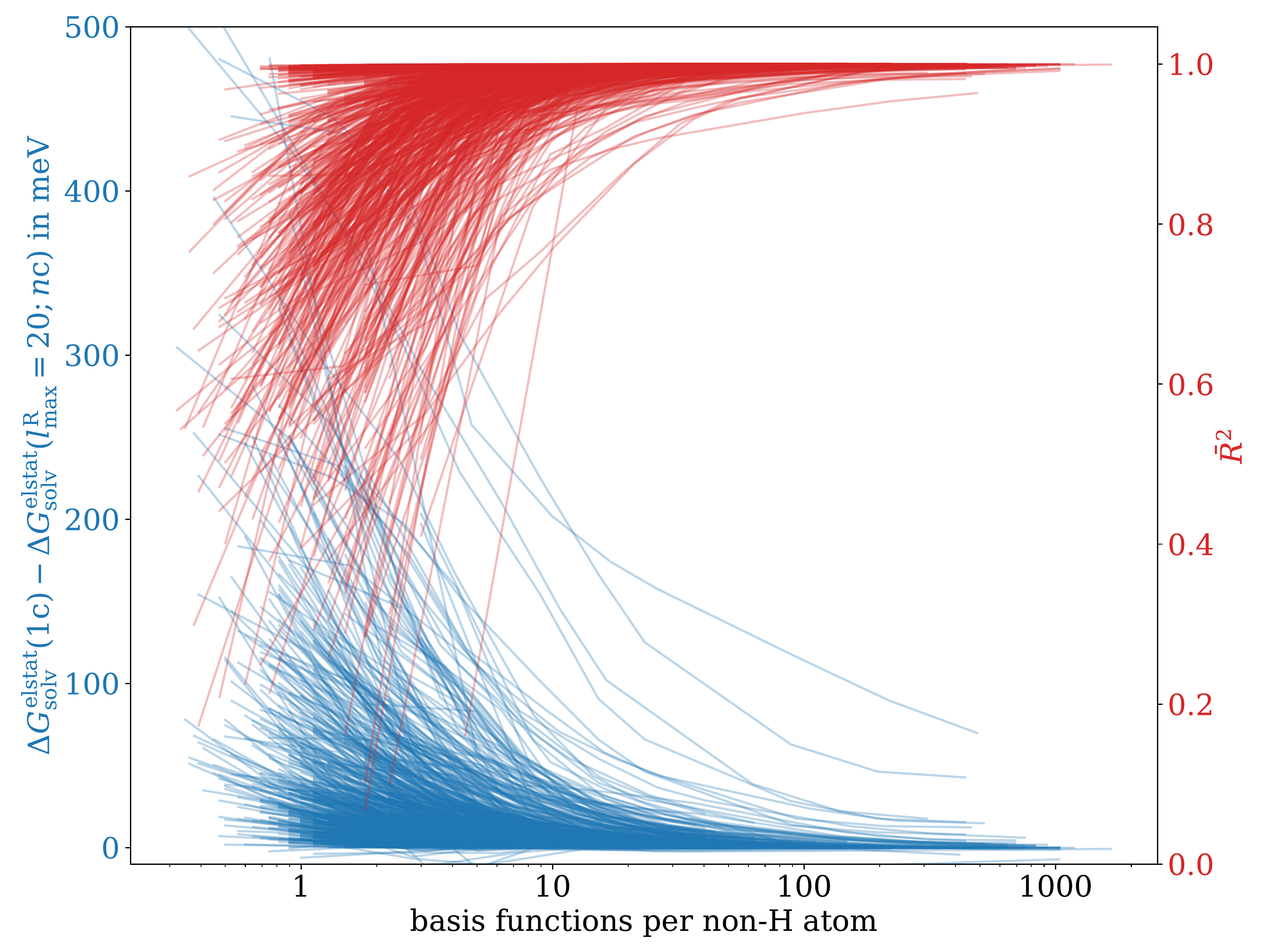}
    \caption{Same as fig. 1, but now showing the convergence with respect to the number of basis functions per non-H atom (on a logarithmic scale) for $\Phi_{\rm R}$ in MPE-1c.\note{opacity of lines increased for better visibility}}
    \label{fig:converge_lmax_rf_nosubcav_bf}
\end{figure}

This raises the question if the error for larger solutes in MPE-1c might simply be due to the basis set being too small. Indeed, already in the original publications\cite{Sinstein,Sinstein_thesis} it was noted that for larger solutes, higher expansion orders than the originally published $l_{\rm max}^{\rm R}=8$ may be necessary. Specifically, some number of basis functions per non-H atom might exist at which $\Delta G_{\rm solv}^{\rm elstat}$ is reliably converged and $\Bar{R}^2\approx 1$. We test this hypothesis by picking all test systems (58 in total) with ${\Bar{R}^2(l_{\rm max}^{\rm R}=20)\leq 0.99}$ and running additional calculations at \change{$l_{\rm max}^{\rm R}=40,60$ and $90$}. The data from \cref{fig:converge_lmax_rf_nosubcav} together with these additional calculations, plotted against the number of basis functions per non-H atom, is shown in \cref{fig:converge_lmax_rf_nosubcav_bf}. 

While a remarkable portion of test systems converges somewhere around 100 basis functions per non-H atom, there are some systems which do not converge to reasonable accuracy even within the enlarged range of $l_{\rm max}^{\rm R}$. It is worth looking at the systems for which the model performs worst in some detail. We identify the systems for which at least one calculation was successfully conducted which fulfills the following two conditions: First, the number of basis functions for $\Phi_{\rm R}$ per non-H atom is at least 81, and second, $\Bar{R}^2<0.98$ and/or the error in $\Delta G_{\rm solv}^{\rm elstat}$ is $>20\,{\rm meV}$. We will show in \cref{sec:convergence} that these conditions do not apply for any of our test systems studied with the new model described in the following sections. 

\begin{figure}[ht!]
\begin{subfigure}{.45\linewidth}
    \centering
    \includegraphics[width=\linewidth]{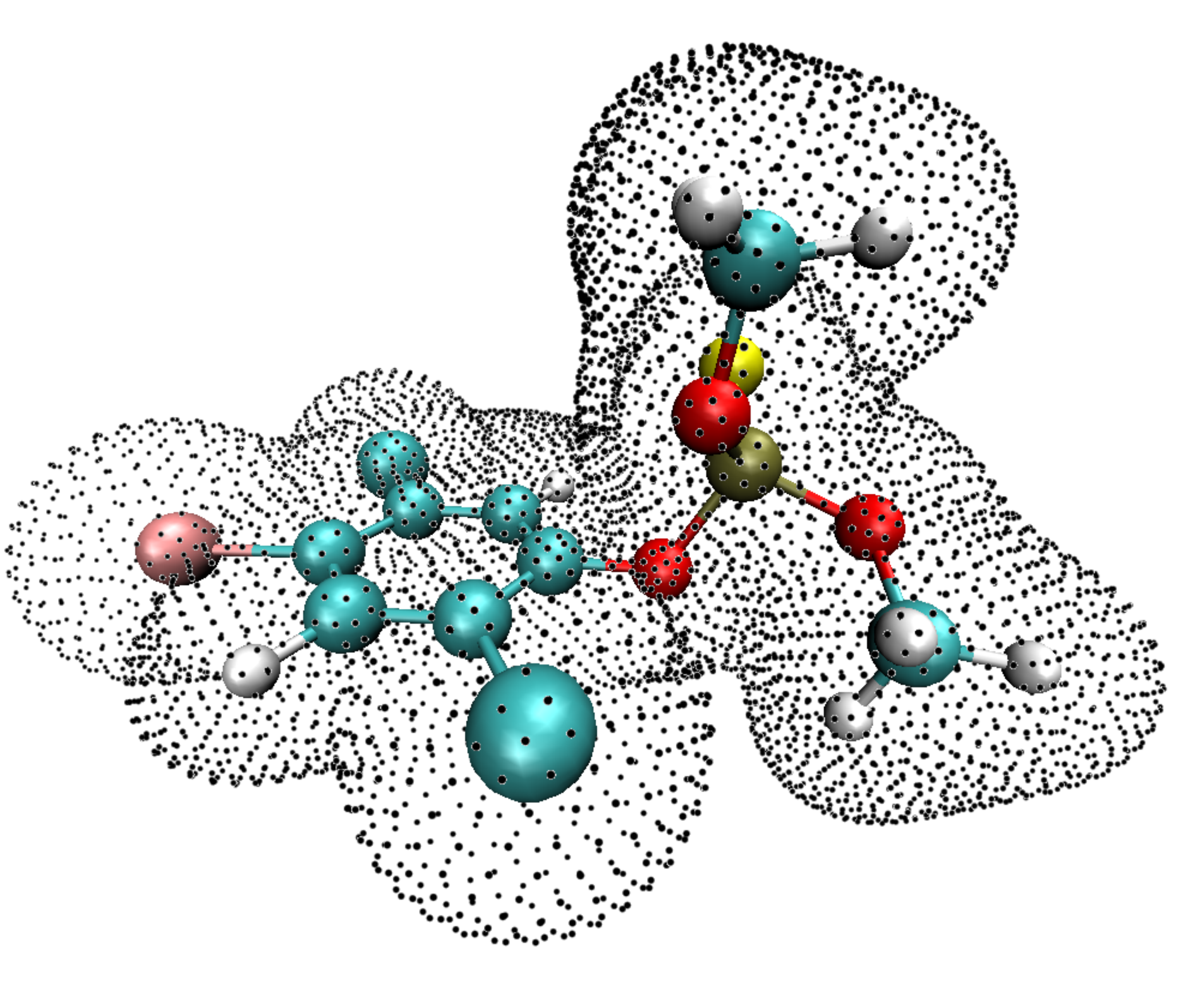}
\end{subfigure}
\begin{subfigure}{.45\linewidth}
    \centering
    \includegraphics[width=\linewidth]{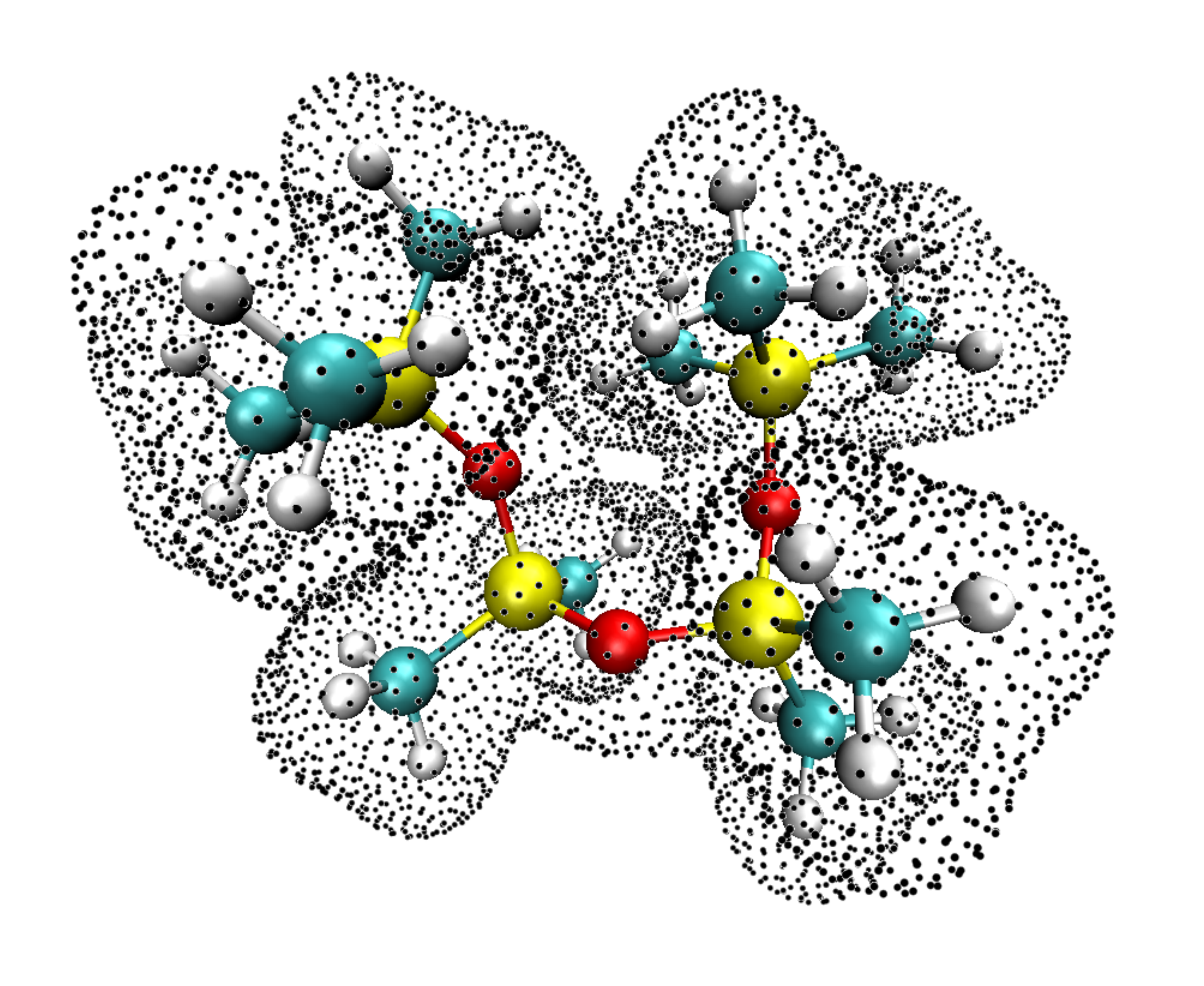}
\end{subfigure} \\
\begin{subfigure}{.45\linewidth}
    \centering
    \includegraphics[width=\linewidth]{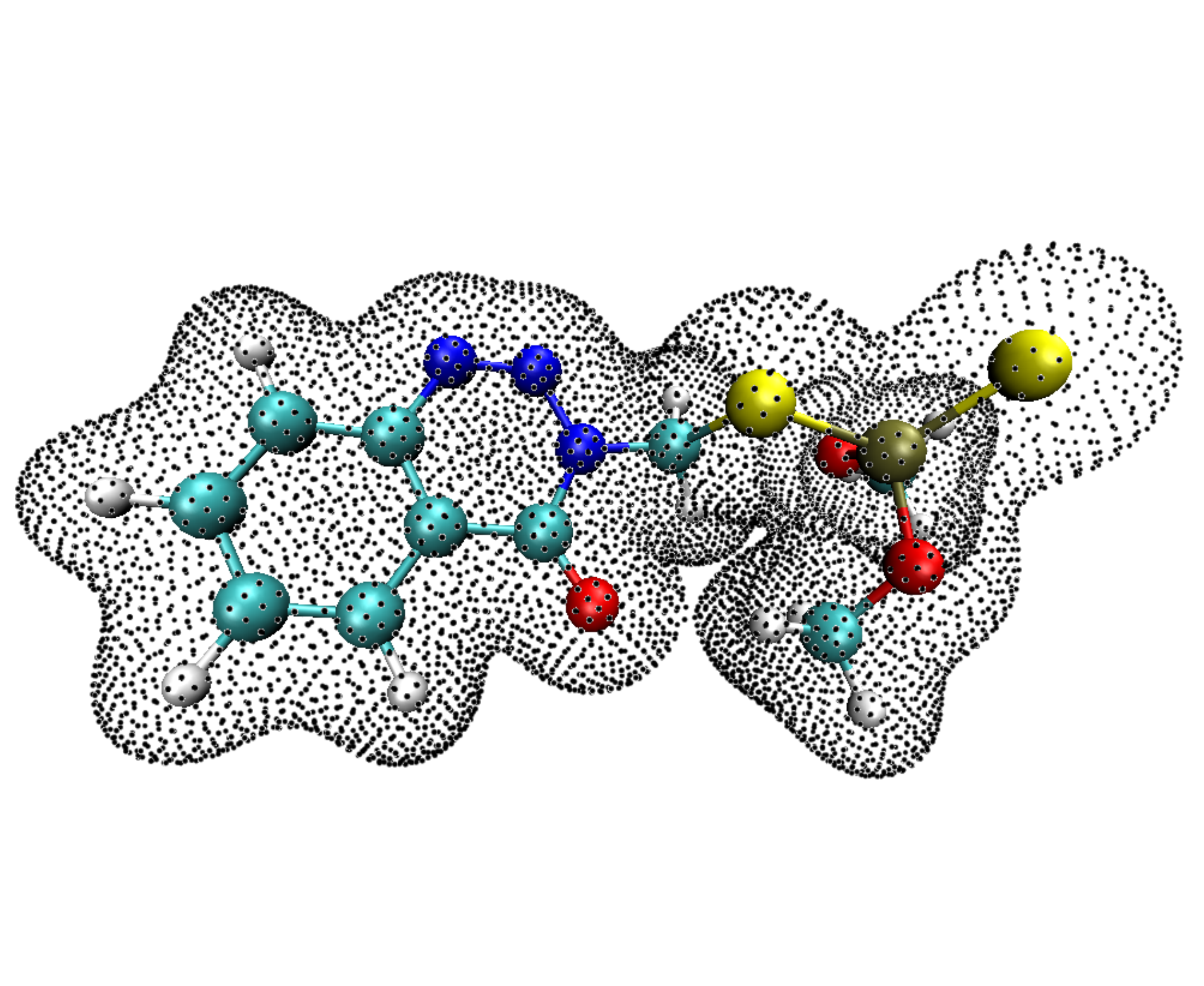}
\end{subfigure}
\begin{subfigure}{.45\linewidth}
    \centering
    \includegraphics[width=\linewidth]{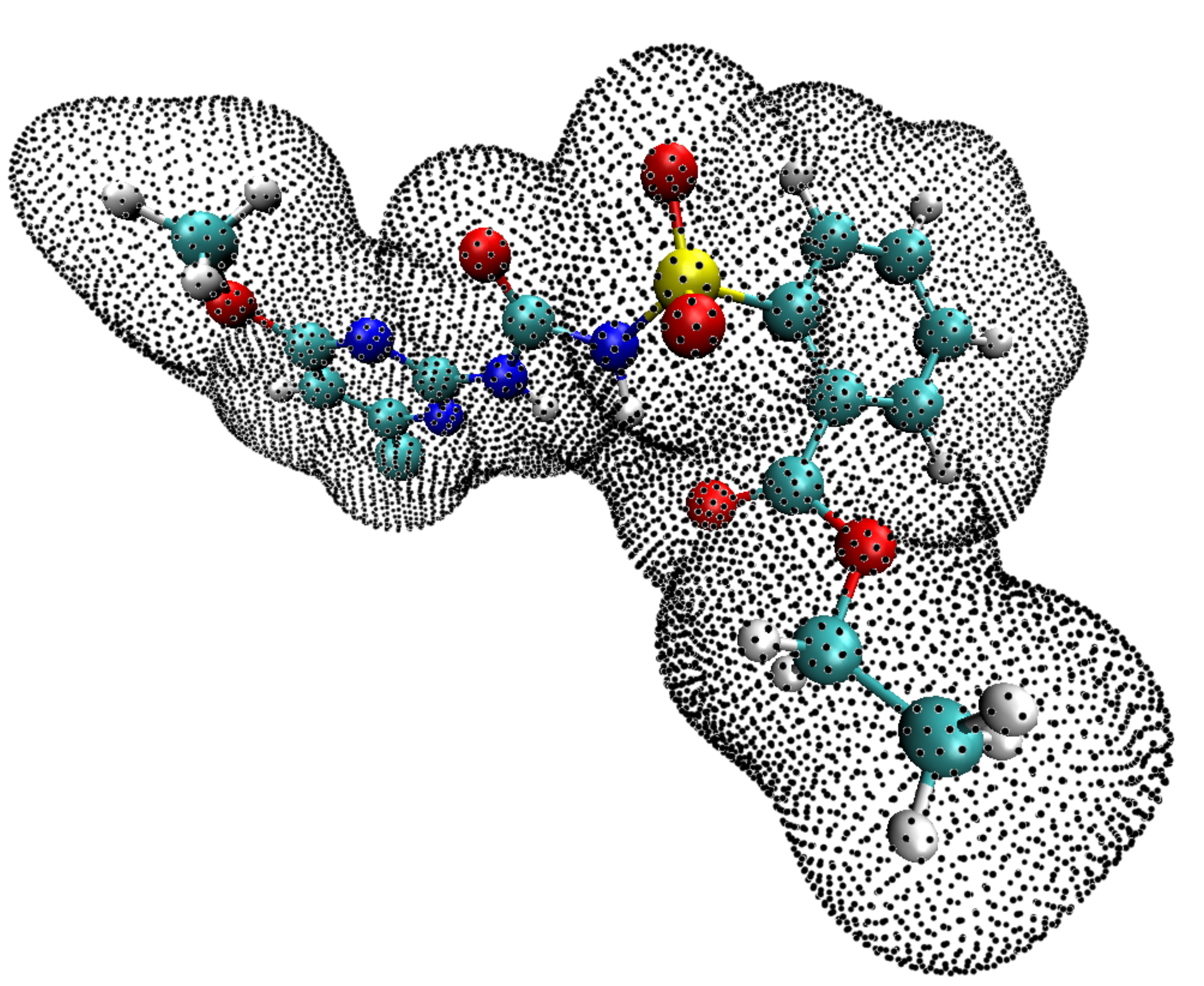}
\end{subfigure} \\
\begin{subfigure}{.45\linewidth}
    \centering
    \includegraphics[width=\linewidth]{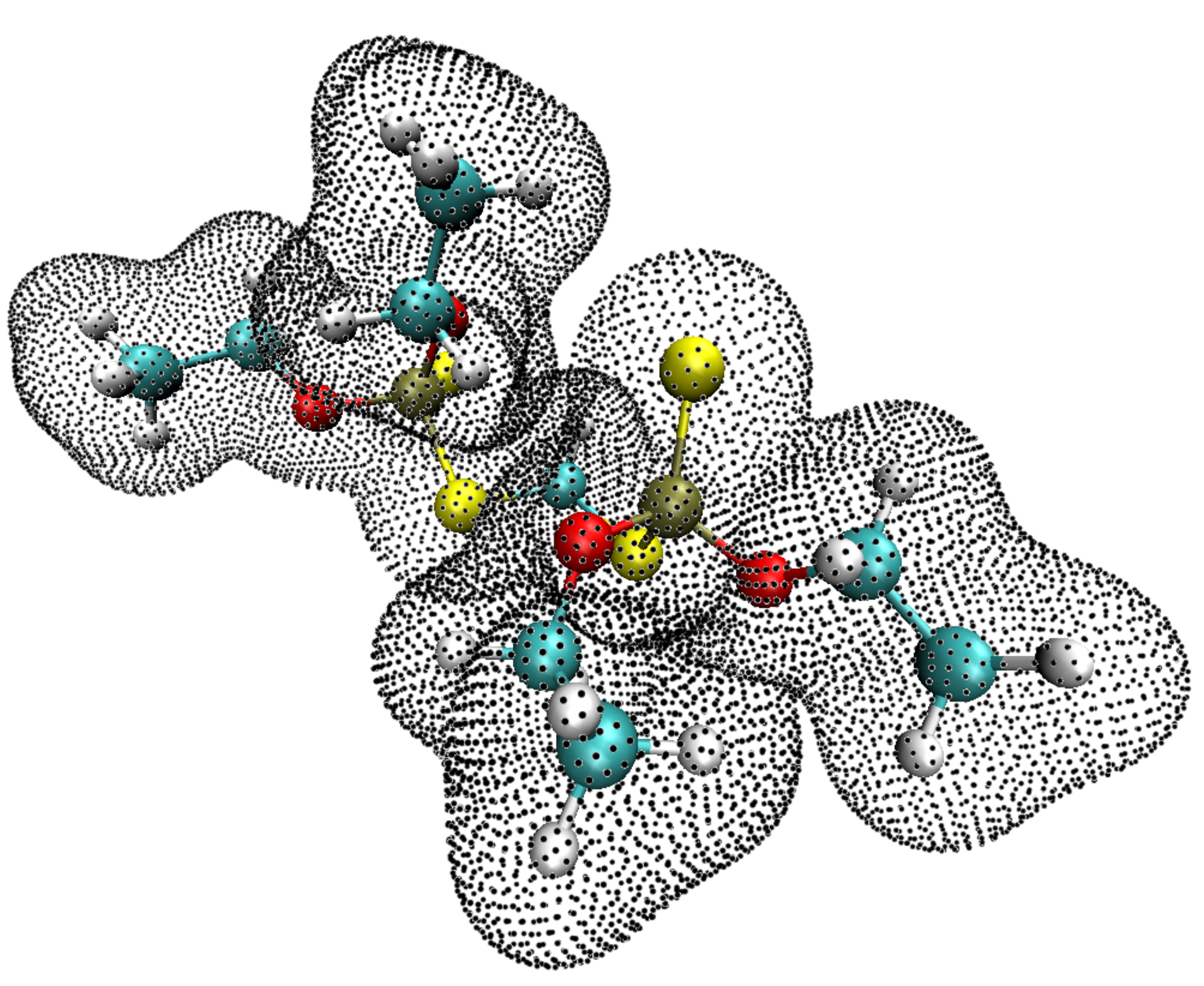}
\end{subfigure}
\begin{subfigure}{.45\linewidth}
    \centering
    \includegraphics[width=\linewidth]{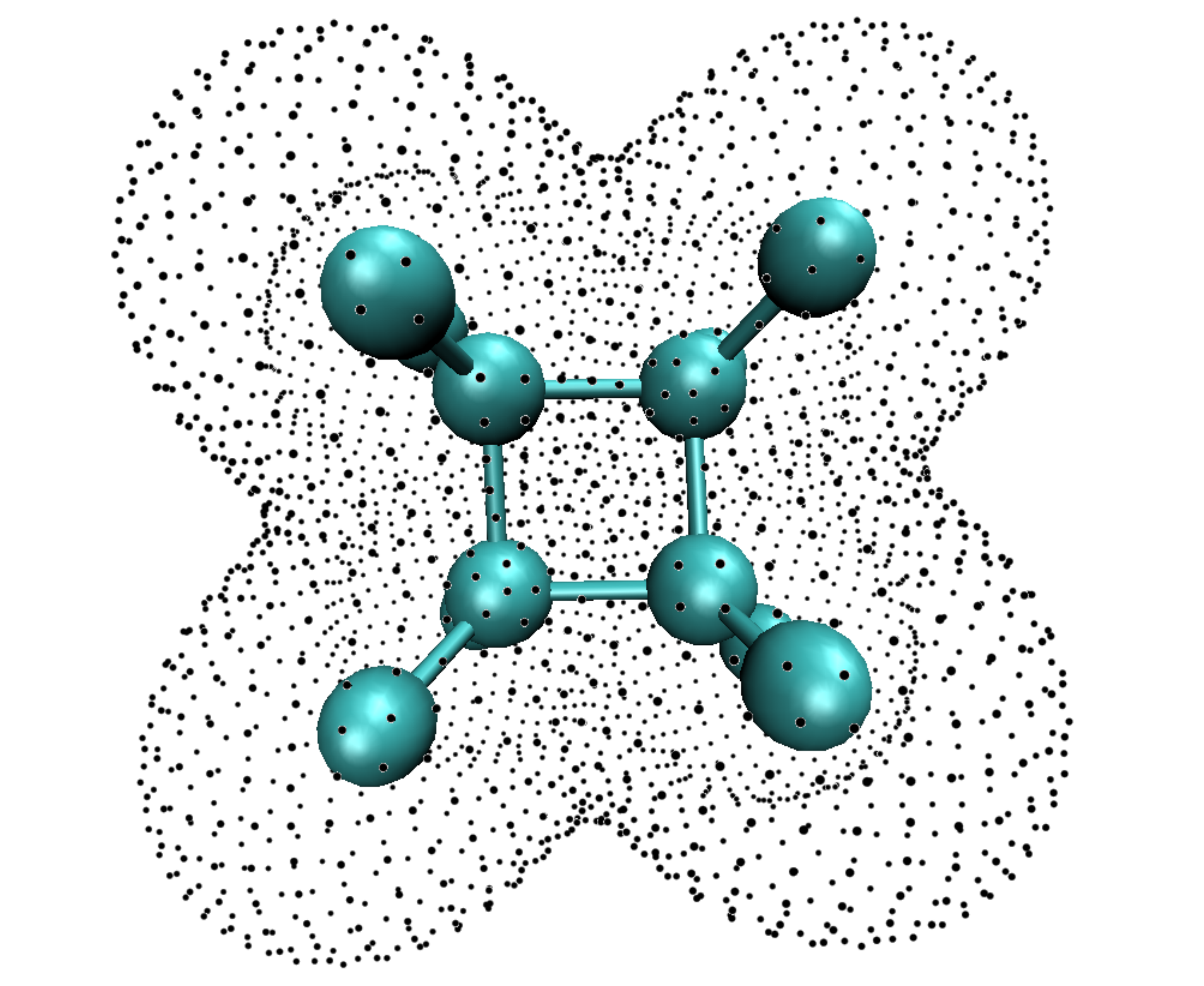}
\end{subfigure}
\caption{Systems described in \cref{tab:mols_unsolvable}, in the same order (row first). Cavity surface sampling density is arbitrary and does not represent the sampling density in the calculations shown in the table.
\label{fig:mols_unsolvable}}
\end{figure}

The six systems identified in this way are shown in \cref{fig:mols_unsolvable} and their parameters and results listed in \cref{tab:mols_unsolvable}. For all these systems, $X_{\rm R}$ deviates far from a convex shape, confirming our explanation from the beginning of this section at least in the sense that when the original method\cite{Sinstein} fails, it is linked to cavities being strongly non-convex. Yet, the inverse conclusion, that non-convex cavities necessarily lead to failure, is not true.

From this data alone it is not entirely clear if the non-convergence is due to incompleteness of the basis for $\Phi_{\rm R}$ alone. As described at the beginning of \cref{sec:original_insuff}, we compare MPE-1c with $l_{\rm max}^{\rm Q}=6$ to a reference with $l_{\rm max}^{\rm Q}=8$. The objective was to assess the performance of MPE-1c as originally published in ref.~\citenum{Sinstein}, where a value of 6 was proposed for use with \textit{tight} integration grids in \texttt{FHI-aims}\cite{Sinstein,aims-paper,aims-Havu2009,aims-Yu2018,aims-Ren2012,aims-Ihrig2015}. We note that the lower expansion order does not automatically mean a smaller basis for $\Phi_{\rm Q}$, because MPE-1c places expansion centers on all solute nuclei, whereas our reference places them only on non-H nuclei, cf.~\cref{sec:problem_definition}. Nonetheless, we verify that the basis for $\Phi_{\rm Q}$ does not cause the observed errors by rerunning the six systems in \cref{tab:mols_unsolvable} with $l_{\rm max}^{\rm Q}=8$, ensuring a larger (or equally large in the case of octafluorocyclobutane) basis for $\Phi_{\rm Q}$ than in the reference. While minor improvements to both $\Bar{R}^2$ and $G_{\rm el}$ are observed, the larger basis for $\Phi_{\rm Q}$ clearly does not solve the issue, confirming that it is, indeed, tied to an incomplete basis for $\Phi_{\rm R}$.

\begin{table}[ht]
    \setlength{\tabcolsep}{0.5em}
    \centering
    \begin{tabular}{l|r|r|r|r|r|r|r}
        solute & $N$ & $\frac{\rho_{\rm iso}}{1\,{\rm me \si{\angstrom}^{-3}}}$ & $\varepsilon$ & $n_{\rm bf, rel}$ & $l_{\rm max}^{\rm Q}$ & $\Bar{R}^2$ & $\frac{\Delta\Delta G_{\rm solv}^{\rm elstat}}{1\,{\rm meV}}$ \\\hline & & & & & & & \\
        0445pho  & 16 &  59.8 &  6.51    & 105.1 & 6 & 0.981 & 23.9 \\
                 &    &       &          &       & 8 & 0.982 & 22.3 \\
        0925dec  & 17 &  82.4 &  3.19    & 98.9  & 6 & 0.939 & 114.4 \\
                 &    &       &          &       & 8 & 0.943 & 110.3 \\
        test1010 & 19 &  94.7 &  4.12    & 88.5  & 6 & 0.983 & 28.4 \\
                 &    &       &          &       & 8 & 0.984 & 26.4 \\
        test1020 & 27 &  30.1 &  5.18    & 137.8 & 6 & 0.986 & 22.3 \\
                 &    &       &          &       & 8 & 0.988 & 20.3 \\
        test1031 & 19 &  64.4 &  346.30  & 88.5  & 6 & 0.975 & 63.0 \\
                 &    &       &          &       & 8 & 0.977 & 58.0 \\
        test2023 & 12 &  49.5 &  6.61    & 140.1 & 6 & 0.975 & 5.2  \\
                 &    &       &          &       & 8 & 0.979 & 3.7
    \end{tabular}
    \caption{Systems with $\Bar{R}^2<0.98$ and/or electrostatic solvation free energy error $\Delta\Delta G_{\rm solv}^{\rm elstat}>20\,{\rm meV}$ in MPE-1c at number $n_{\rm bf,rel}\geq 81$ of basis functions for $\Phi_{\rm R}$ per non-H atom. $N$ is the number of non-H atoms. Solutes are identified by their entry number in the Minnesota Solvation Database\cite{MinSol}. For each system, of all calculations fulfilling these conditions, only the one with the lowest $l_{\rm max}^{\rm R}$ is shown.}
    \label{tab:mols_unsolvable}
\end{table}

Overall, six systems with unsatisfactory solutions to the electrostatic problem may not sound like a lot. All of these system do, however, lie in the sparsely sampled upper end of our test set regarding molecular size (cf.~\cref{fig:hist_n_nonH}). Furthermore, their $\rho_{\rm iso}$ all lie in the upper end of the sampled range, where we generally expect more complicated cavity shapes. Looking only at systems with 16 or more non-H atoms and $\rho_{\rm iso}\geq 30\,{\rm me \si{\angstrom}^{-3}}$, we find that of only 12 such systems in our test set, 5 failed in the above described sense. Thus, even after weighting the basis set size by the number of non-H atoms, larger molecules in combination with high isodensities still show a much worse convergence (or possibly none at all) with expansion order than the smaller molecules or systems with lower isodensities.

One may argue that these errors are due to small voids in the solutes. Such voids can be assigned to the dielectric medium $X_{\rm Q}$ following only the simplistic cavity definition in \cref{eq:def_cavity_orig,eq:def_dielectric_orig}. If they are smaller than a solvent molecule, the resulting model is obviously unphysical. Indeed, such voids are observed in some of the problematic cases described above, probably most evidently in decamethyltetrasiloxane (entry 0925dec in the Minnesota Solvation Database\cite{MinSol}, top right in \cref{fig:mols_unsolvable}). There exist solutions to this issue, e.g. using so-called solvent-aware interfaces.\cite{solvent_aware_interface} Implementing such a method is, however, beyond the scope of the present work.

For the time being, we state that although the convergence with expansion order may improve with a more physically motivated cavity definition, the issues presented in this section still persist. First, we observe the issue also in some systems without obviously unphysical cavity shapes, particularly chlorimuron-ethyl (test1020 in the Minnesota Solvation Database\cite{MinSol}) and octafluorocyclobutane (test2023 in the Minnesota Solvation Database\cite{MinSol}) (center right and bottom right in \cref{fig:mols_unsolvable}). Second, our model is agnostic of the solvent's molecular size and ideally the solution ansatz should work independent of cavity choice. On a similar notion, when going to larger solutes than our test set, e.g.~biomolecules or catalytic surfaces, the issue of non-convex cavities will inevitably emerge again.

We conclude that for small molecules, MPE-1c exhibits remarkable robustness once $l_{\rm max}^{\rm R}$ is chosen based on the number of basis functions per non-H atom. For a significant portion of the larger solutes, however, the basis expansion converges slowly or not at all. In the following sections we therefore develop a modified MPE model which solves this issue, aiming at reliable convergence with expansion order.

\section{\label{sec:subcav}Subcavity method}

As a remedy for the issues discussed above, we modify the MPE method\cite{Sinstein}, putting particular emphasis on the solution of the electrostatic problem in arbitrarily shaped cavities. The main idea of our method, which we call MPE-$n$c, is to formally not solve the problem for one single cavity $X_{\rm R}$, but for multiple small `subcavities' $X_{\text{R},K}$. \change{A minimal example for the user input required to use our method in \aims is provided in the SI.}

\subsection{\label{sec:problem_definition}Separation into subcavities}

\begin{figure}[ht]
    \centering
    \includegraphics[width=\linewidth]{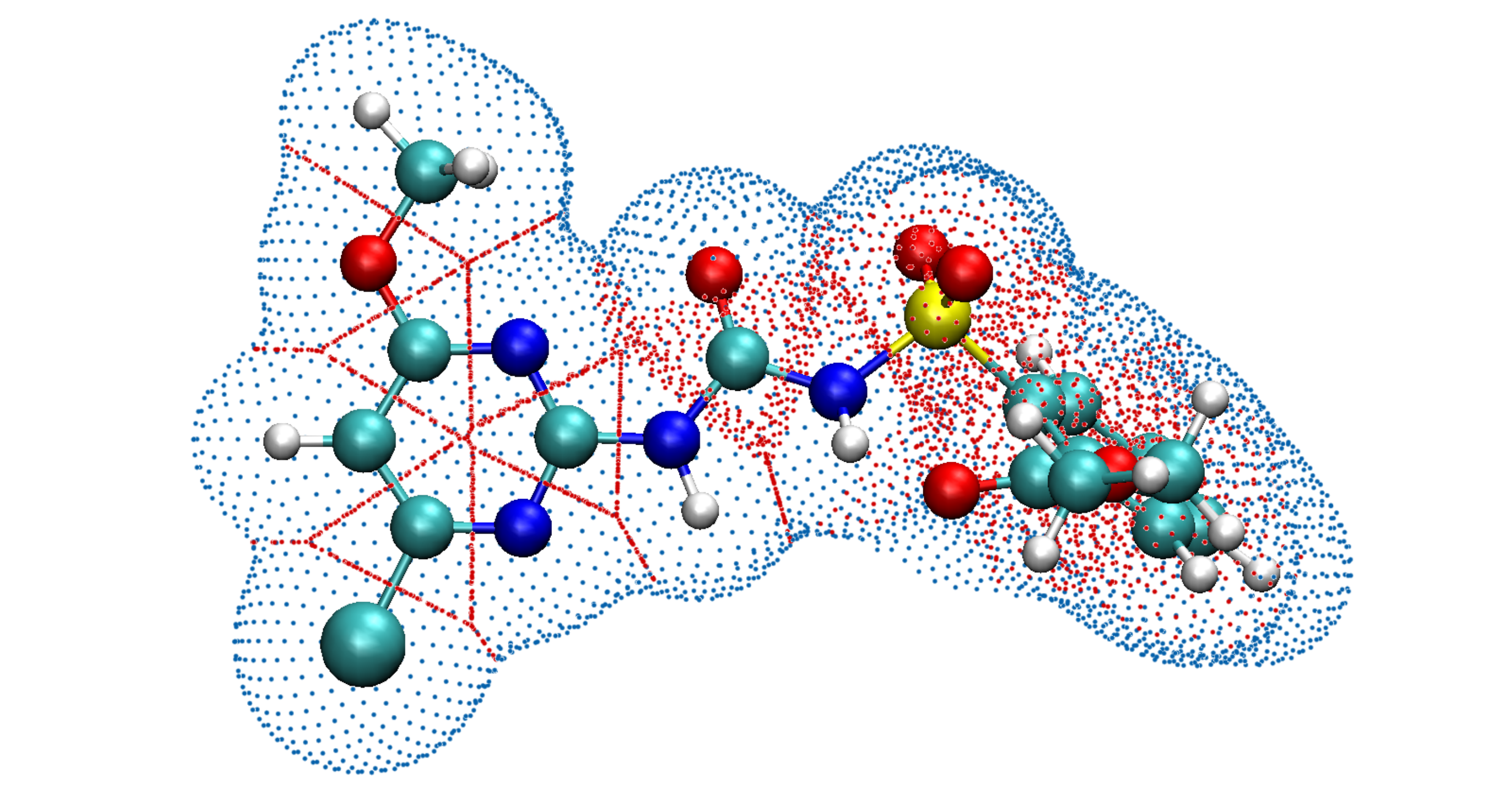}
    \caption{Orthographic view of chlorimuron-ethyl (entry test1020 in Minnesota Solvation Database\cite{MinSol}, cf.~\cref{fig:mols_unsolvable}, center right) with discretized cavity surface (blue) and subcavity-subcavity boundaries (red). Interfaces cut along the plane of the heteroaromatic ring.}
    \label{fig:orthographic}
\end{figure}

We separate $X_\text{R}$ into subcavities $X_{\text{R},K}$ around multiple centers $\mathbf{r}_K$.
\begin{equation}
    X_{\text{R},K} = \{\mathbf{r} : \mathbf{r} \in X_\text{R}, \|\mathbf{r}-\mathbf{r}_K\| < \|\mathbf{r}-\mathbf{r}_{K^\prime}\| \, \forall K^\prime\neq K \} \quad . \label{eq:subcavities} 
\end{equation}
In principle, fast convergence of a series expansion in $\{\mathcal{R}_m^l\}_K$ in these subcavities is to be expected for any choice of $\mathbf{r}_K$, as long as the cavity can approximately be described as a superposition of spheres around these centers. The straightforward choice is to use the positions of all\note{removed} non-H nuclei of the solute. We use this set of $\mathbf{r}_K$ throughout the rest of this work. An example of such a partition is shown in \cref{fig:orthographic}. \change{The basic idea of decomposing an implicit solvent cavity into atom-centered domains and solving the individual electrostatic problems in a mutually consistent way has previously been applied to COSMO.\cite{ddcosmo1,ddcosmo2,ddcosmo3} Here, we discuss the application of this idea to the MPE model.}

$\Phi_\text{MPE}$ is\note{removed} defined piece-wise in each of these subcavities
\begin{equation}
    \Phi_\text{MPE}(\mathbf{r}\in X_{\text{R},K}) = \Phi_{\text{R},K}(\mathbf{r})
\end{equation}
The reaction field $\Phi_{\text{R},K}$ in the respective subcavities can be expanded according to \cref{eq:expansion_R_general},\cite{Sinstein}
\begin{equation}\label{eq:expansion_phir}
	\Phi_{\text{R},K}(\mathbf{r}) = \sum_{l=0}^{l^\text{R}_{\text{max}}} \sum_{m=-l}^l R_K^{(l,m)} \mathcal{R}_m^l(\mathbf{r}-\mathbf{r}_K) \quad .
\end{equation}
Similarly, the external potential $\Phi_\text{Q}$ can be expanded according to \cref{eq:expansion_Q_general},\cite{Sinstein}
\begin{equation}\label{eq:expansion_phiq}
	\Phi_\text{Q}(\mathbf{r}) = \sum_{J=1}^{J_\text{max}} \sum_{l=0}^{l^\text{Q}_{\text{max}}} \sum_{m=-l}^l Q_J^{(l,m)} \mathcal{I}_m^l(\mathbf{r}-\mathbf{r}_J) \quad .
\end{equation}
While the choice of expansion centers is, once again, arbitrary within certain limits, we use the same set of expansion centers $\{\mathbf{r}_J\}=\{\mathbf{r}_K\}$ as for the reaction field in the subcavities, i.e. the non-H nuclear positions of the solute. This is different from the original version\cite{Sinstein} of the method, which placed expansion centers for $\Phi_{\rm Q}$ also on hydrogen cores. We show in \cref{sec:convergence} that $\Phi_{\rm Q}$ still converges using this smaller basis set, saving computational resources.\note{removed} Similarly, it is in principle possible to use different expansion orders for different centers. Since all our centers are, however, qualitatively similar, there is no need to do this. Therefore, we use only two different expansion orders - one for all $\Phi_{\text{R},K}$ and one for $\Phi_\text{Q}$.

\change{In complete analogy to MPE-1c, the boundaries $B_{ij}$ are discretized into a finite set of points, aiming at some target degree of determination $d_{\rm det}$ of the overall linear system \cref{eq:SLE}. Additionally, the algorithm tries to sample all boundaries with approximately the same density of points. The exact discretization procedure for the subcavity-subcavity boundaries $B_{\text{R}K, \text{R}K^\prime}$, and a modification to the discretization of the subcavity surfaces $B_{\text{Q,R}K}$ with respect to the original method\cite{Sinstein} are described in the SI. There, we also describe how the uniform coordinate scaling used to improve matrix conditioning\cite{Sinstein_thesis} is adjusted to MPE-$n$c.

The matrix $\mathbf{A}$ in the central linear system \cref{eq:SLE} becomes sparse with this piecewise multipole expansion, because the columns corresponding to the basis functions for one region $X_i$ have non-zero elements only in the rows corresponding to points on boundaries $B_{ij}$ which are in contact with $X_i$.\cite{Sinstein_thesis} The exact structure of $\mathbf{A}$, as well as an implementation of the iterative LSQR\cite{lsqr} solver that exploits this sparsity, are elaborated in the SI. Alternatively, one can use one of the direct solution algorithms reported in ref.~\citenum{Sinstein}. First, QR factorization of $\mathbf{A}$ is performed. For well-conditioned systems, \cref{eq:SLE} can afterwards be solved straightforwardly. We call this procedure the `QR' solver henceforth. If the resulting matrix $\mathbf{R}$ is rank deficient, however, it can be further factorized in a singular value decomposition (SVD), and a regularized solution can be obtained by applying a cutoff in the inversion of the singular values. We call the latter option the `QR+SVD' solver henceforth. Note that the QR and QR+SVD solvers do not exploit sparsity. The implications of this issue are discussed in \cref{sec:performance}.}

\note{sections moved to SI}

\subsection{Charge conservation \label{sec:charge}}

We take the opportunity of generalizing MPE implicit solvation to address another, minor issue which is not directly related to those described in \cref{sec:original_insuff,sec:phyical_or_technical}.

From Gauss's law it follows that
\begin{equation}\label{eq:charge_conv}
    \oint_{\mathbf{A}\subset X_{\rm Q}} \nabla \Phi_{\rm Q}\, d\mathbf{A} = 0
\end{equation}
for arbitrary closed surfaces $\mathbf{A}$ lying entirely in $X_{\rm Q}$. In principle, this should be fulfilled automatically, if \cref{eq:continuity_boundary_general_field} is fulfilled exactly on all boundaries $\change{B}_{ij}$. Our method is, however, approximate by nature and boundary conditions are fulfilled only in a least-squares sense. Usually, this does not lead to any severe or systematic errors. \change{In the case of \cref{eq:charge_conv}, however, the error formally amounts to a charge, altering the effective charge $\oint_{\mathbf{A}\subset X_{\rm Q}} \mathbf{D}\, d\mathbf{A}$ of the system in solution. We note that this is unrelated to the so-called outlying charge error concerning explicit electron density, which was already discussed and resolved in refs.~\citenum{Sinstein,Sinstein_thesis}. The error discussed here concerns solely the implicit potential. The} existence of a spurious charge alone may, in some cases, be enough to introduce a systematic error in the electrostatic energy. It is therefore worth enforcing \cref{eq:charge_conv} not in a least-squares fashion, but exactly (up to numeric precision). \change{The technical details on the constrained solution algorithm are explained in the SI.}

\note{moved to SI}

\section{\label{sec:results}Results and discussion}

With the details of the method now in place we turn to the performance of MPE-$n$c and provide numeric parameters for practical applications. \note{removed}

\subsection{Convergence with numeric parameters \label{sec:convergence}}

\begin{figure}[ht]
    \centering
    \includegraphics[width=\linewidth]{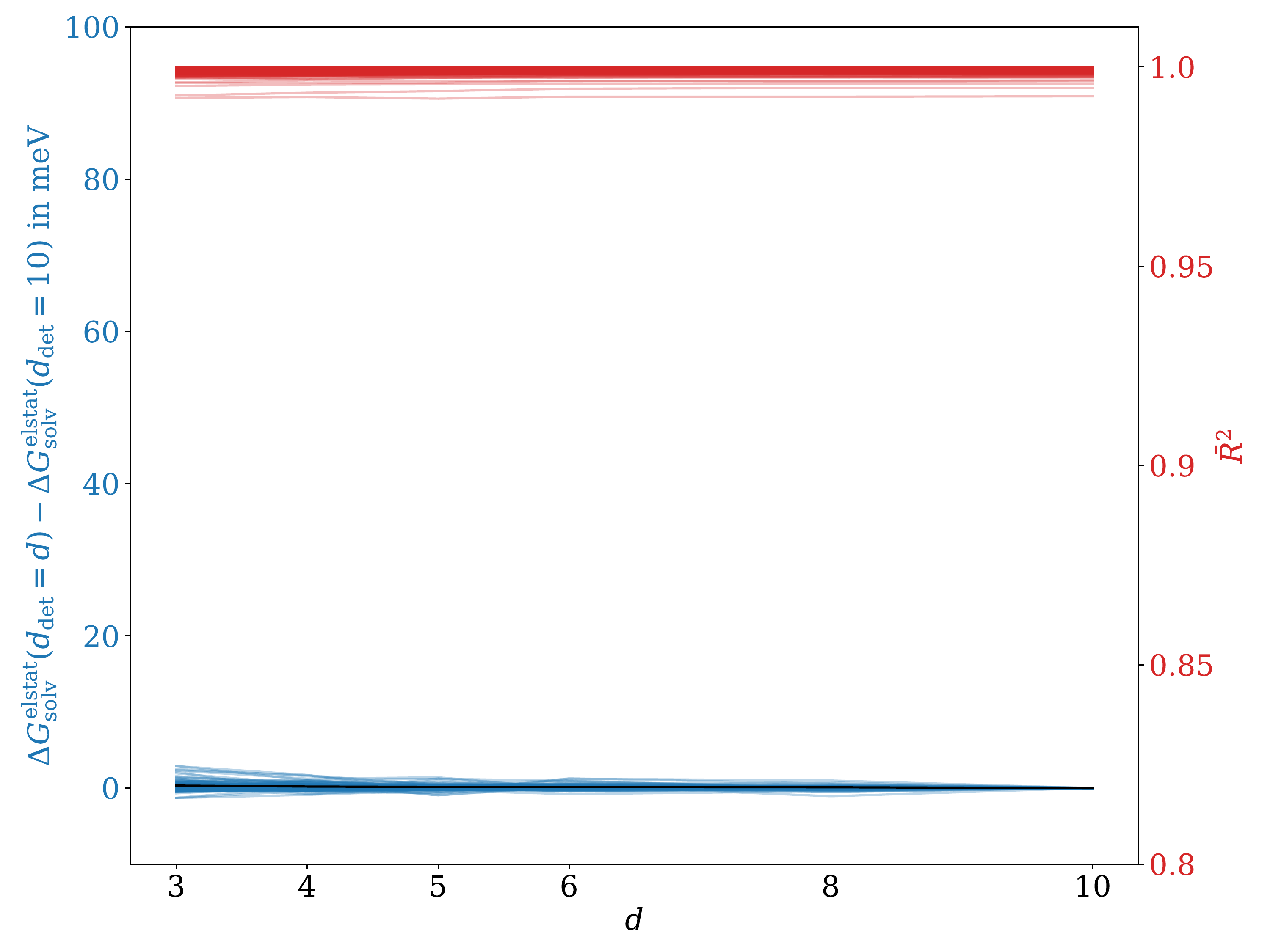}
    \caption{Convergence of the electrostatic contribution to the solvation free energy $G_{\rm solv}^{\rm elstat}$ (blue) and adjusted coefficient of determination $\Bar{R}^2$ (red) of the discretized boundary conditions \cref{eq:continuity_boundary_general_pot,eq:continuity_boundary_general_field} with target degree of determination $d_{\rm det}$ of the SLE\note{removed}.\note{opacity of lines increased for better visibility}}
    \label{fig:converge_dod}
\end{figure}

\begin{figure}[ht]
    \centering
    \includegraphics[width=\linewidth]{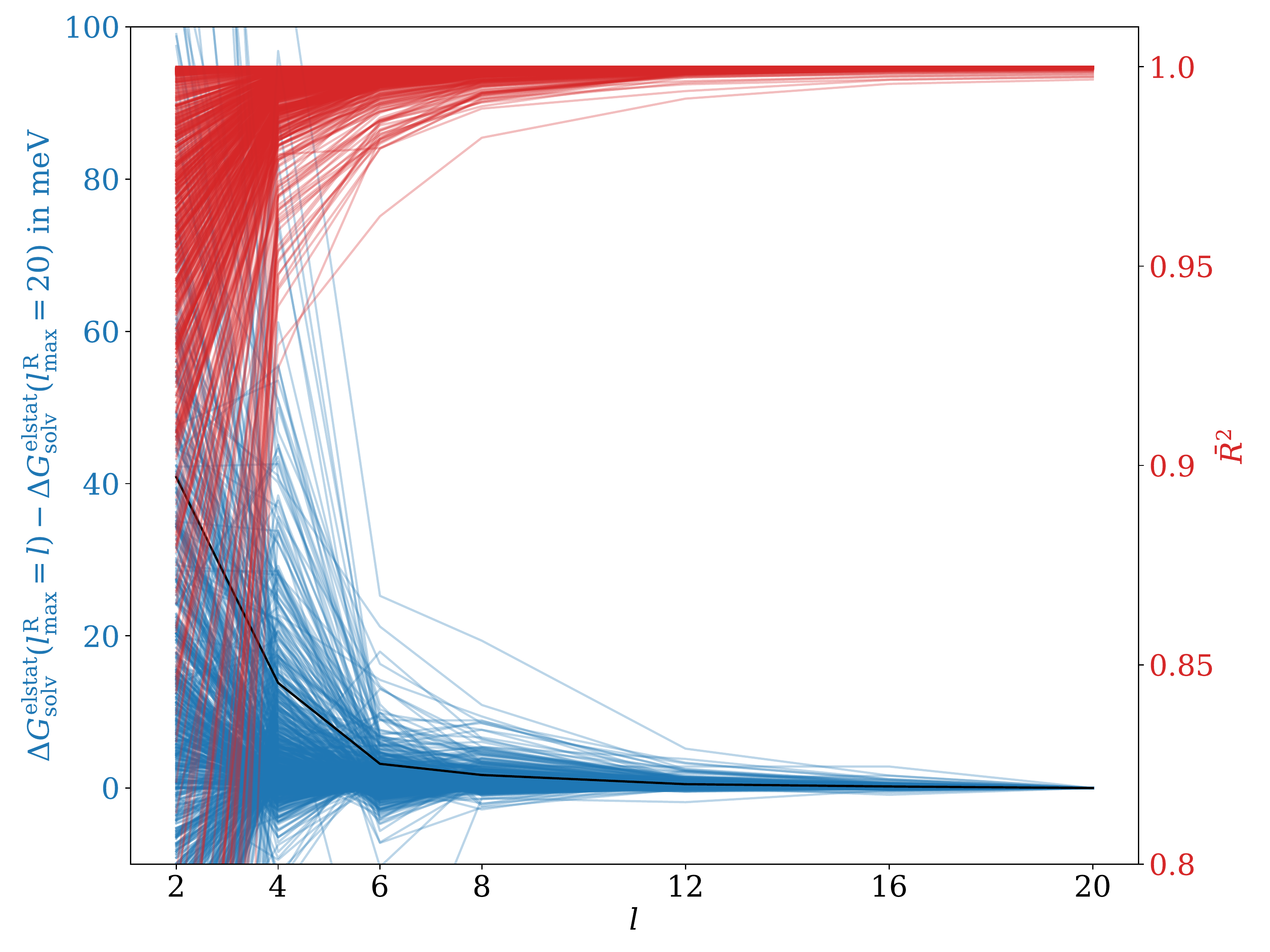}
    \caption{Same as fig. 8, but showing convergence with expansion order $l_{\rm max}^{\rm R}$ of the MPE potential inside the subcavities.\note{opacity of lines increased for better visibility}}
    \label{fig:converge_lmax_rf}
\end{figure}

\begin{figure}[ht]
    \centering
    \includegraphics[width=\linewidth]{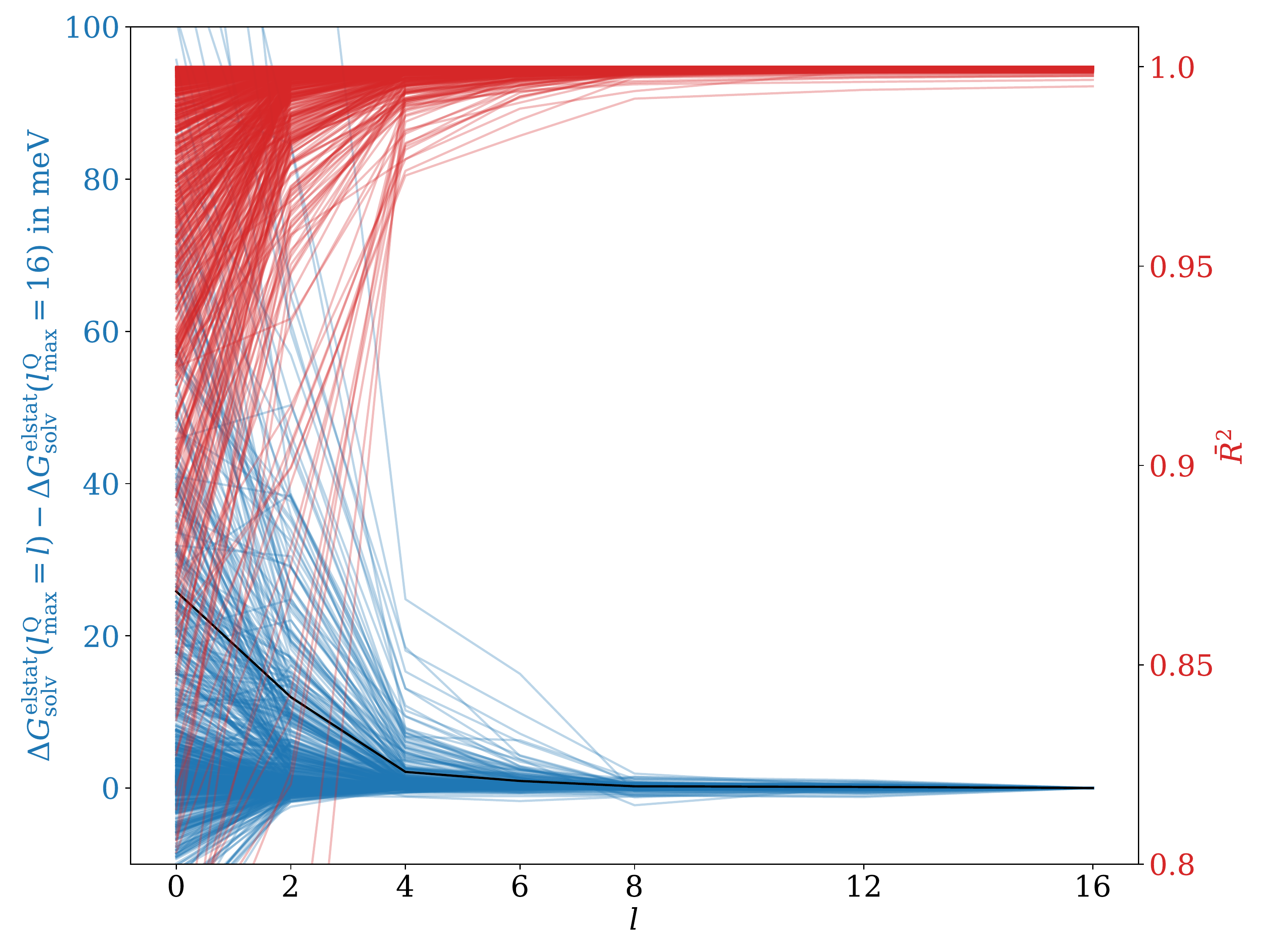}
    \caption{Same as fig. 8, but showing convergence with expansion order $l_{\rm max}^{\rm Q}$ of the MPE potential outside the cavity.\note{opacity of lines increased for better visibility}}
    \label{fig:converge_lmax_ep}
 \end{figure}

We test the convergence of $\Delta G_{\rm solv}^{\rm elstat}(n{\rm c})$ and $\Bar{R}^2$ with expansion orders $l^{\rm R/Q}_{\rm max}$ and target degree of determination $d_{\rm det}$. For each of the three parameters, values are sampled on respective grids \change{$l^{\rm R}_{\rm max}=2,4,6,8,12,16,20$}, \change{$l^{\rm Q}_{\rm max}= 0,2,4,6,8,12,16$} and \change{$d_{\rm det}=3,4,5,6,8,10$}. We vary each of the three convergence parameters separately and set the respective other two to their `\textit{really tight}' defaults, as discussed below. These are $l^{\rm R}_{\rm max}=12$, $l^{\rm Q}_{\rm max}=8$ and $d_{\rm det}=5$. These values were determined from preparatory runs, which were then confirmed in the here reported results. We therefore do not give an extra account of these initial runs, but rather focus on the full results summarized in \cref{fig:converge_lmax_rf,fig:converge_lmax_ep,fig:converge_dod}. In order to illustrate the convergence with these values, we use $\Delta G_{\rm solv}^{\rm elstat}(n{\rm c})$ at $l^{\rm R}_{\rm max}= 20$ for the $l^{\rm R}_{\rm max}$ convergence study, at $l^{\rm Q}_{\rm max}=16$ for the $l^{\rm Q}_{\rm max}$ convergence study and at $d_{\rm det}=10$ for the $d_{\rm det}$ convergence study as a converged reference, respectively.

As depicted in \cref{fig:converge_dod}, the target degree of determination $d_{\rm det}$ appears to have no noticeable influence. The RMSE with respect to the reference value is consistently smaller than $1\,{\rm meV}$ and shows little dependence on $d_{\rm det}$. Furthermore, no significant outliers are observed. Nonetheless, to avoid parts of the SLE becoming underdetermined, $d_{\rm det}$ should not be chosen too small.\cite{Sinstein_thesis} Indeed, the ratio of non-zero rows to columns for individual potentials $\Phi_i$ may be smaller than $2d_{\rm det}$, as regions $X_i$ have different numbers of basis functions and discretized points in adjacent interfaces $\change{B}_{ij}$. Choosing a safe, large enough $d_{\rm det}$ does not increase the computational cost too dramatically, as it enters the matrix size only linearly. All in all, $d_{\rm det}$ can be considered a fairly uncritical parameter and can be chosen higher or lower where needed. For the rest of the present work, we use the previously established\cite{Sinstein} value of $d_{\rm det}=5$, which yielded an RMSE of $0.15\,{\rm meV}$ with respect to the reference value in our convergence study.

\begin{figure}[ht]
\begin{subfigure}{.45\linewidth}
    \centering
    \includegraphics[width=\linewidth]{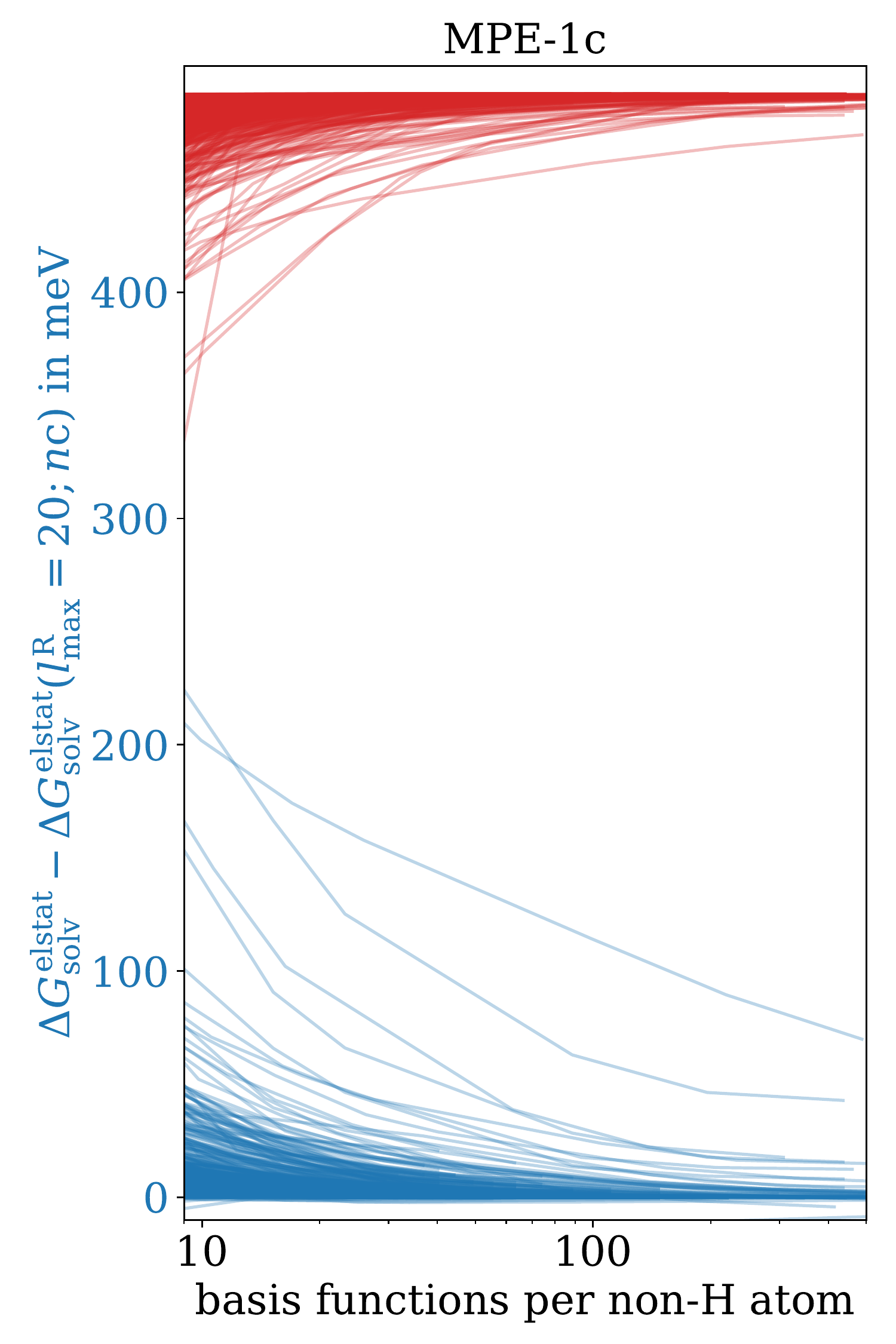}
\end{subfigure}
\begin{subfigure}{.45\linewidth}
    \centering
    \includegraphics[width=\linewidth]{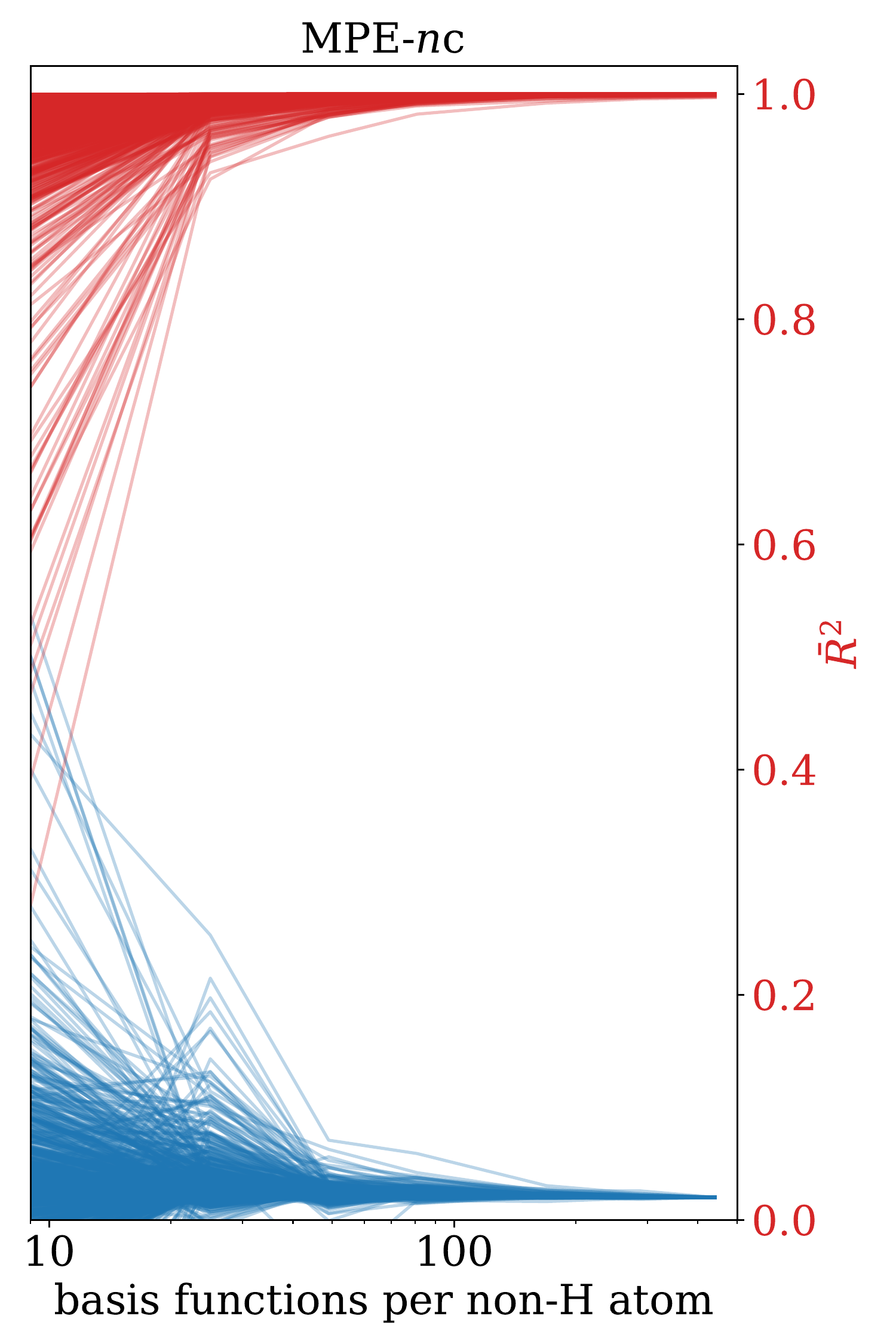}
\end{subfigure}
\caption{Convergence of the electrostatic contribution to the solvation free energy $\Delta G_{\rm solv}^{\rm elstat}$ (blue) and adjusted coefficient of determination $\Bar{R}^2$ (red) of the discretized boundary conditions \cref{eq:continuity_boundary_general_pot,eq:continuity_boundary_general_field} with number of basis functions per non-H atom (on a logarithmic scale) for $\Phi_{\rm R}$ in \change{the original} (left) and \change{our modified method} (right).\note{opacity of lines increased for better visibility, labels moved from caption to figure for clarity}
\label{fig:convergence_compare}}
\end{figure}

A more significant convergence behavior of $\Delta G_{\rm solv}^{\rm elstat}$ is instead obtained  with increasing expansion orders $l_{\rm max}^{\rm R/Q}$ as depicted in \cref{fig:converge_lmax_rf,fig:converge_lmax_ep}. In both cases, no noticeable changes in $\Delta G_{\rm solv}^{\rm elstat}$ occur at the upper end of the range of sampled values, indicating that the solution to the electrostatic problem is then, indeed, converged. This is further supported by $\Bar{R}^2$ of the discretized boundary conditions \cref{eq:continuity_boundary_general_pot,eq:continuity_boundary_general_field} converging to $1$. MPE-$n$c shows convergence for all of the test cases, even those that failed to converge with MPE-1c. To illustrate the advantage of MPE-$n$c over MPE-1c we plot in \cref{fig:convergence_compare} the convergence with overall basis set size of $\Phi_{\rm R}$. Contrary to MPE-1c, the new subcavity method 
starts to converge for all geometries at around 50 basis functions per non-H atom.

Similarly to the single cavity case, solvation free energies at too low expansion orders are typically higher than the reference values, indicating a systematic underestimation of the electrostatic solute-solvent interaction at these expansion orders. The error is, however, not strictly variational as overestimation can be observed occasionally.

Assuming that the errors introduced by the individual parameters are uncorrelated, we can estimate the total root mean-square error as
\begin{equation}
\begin{split}
    &{\rm RMSE}_{\rm est}(l_{\rm max}^{\rm R},l_{\rm max}^{\rm Q},d_{\rm det}) =\\
    &\sqrt{{\rm RMSE}(l_{\rm max}^{\rm R})^2+{\rm RMSE}(l_{\rm max}^{\rm Q})^2+{\rm RMSE}(d_{\rm det})^2} \quad ,
\end{split}
\end{equation}
where ${\rm RMSE}(x)$ is the root mean-square error of $\Delta G_{\rm solv}^{\rm elstat}$ at the respective value of $x$ with respect to $\Delta G_{\rm solv}^{\rm elstat}$ at the largest tested  value of $x$. Given that the error with respect to $d_{\rm det}$ stays mostly constant we omit if from further consideration. With respect to the other two error components, though, we can use ${\rm RMSE}_{\rm est}$ to estimate the expected errors for given combinations of $l_{\rm max}^{\rm R},l_{\rm max}^{\rm Q}$. Given that the computational cost of the method rises steeply with increased MPE expansion order it useful to define default values for different levels of expected accuracy.
In analogy to the electronic basis sets of the \aims code itself we therefore define three convergence levels   
\begin{itemize}
    \item `\textit{really tight}': $l^{\rm R}_{\rm max}=12$, $l^{\rm Q}_{\rm max}=8$  \\
     solver: QR+SVD \\
     ${{\rm RMSE}_{\rm est}=0.58\,{\rm meV}}$ \\ ${{\rm RMSE}(l^{\rm R}_{\rm max})=0.51\,{\rm meV}}$, ${{\rm RMSE}(l^{\rm Q}_{\rm max})=0.23\,{\rm meV}}$
    \item `\textit{tight}': $l^{\rm R}_{\rm max}=8$, $l^{\rm Q}_{\rm max}=6$ \\
    solver: any \\
    ${{\rm RMSE}_{\rm est}=2.0\,{\rm meV}}$ \\ ${{\rm RMSE}(l^{\rm R}_{\rm max})=1.7\,{\rm meV}}$, ${{\rm RMSE}(l^{\rm Q}_{\rm max})=0.93\,{\rm meV}}$ 
    \item `\textit{light}': $l^{\rm R}_{\rm max}=6$, $l^{\rm Q}_{\rm max}=4$ \\
    solver: any \\
    ${{\rm RMSE}_{\rm est}=3.8\,{\rm meV}}$ \\ ${{\rm RMSE}(l^{\rm R}_{\rm max})=3.2\,{\rm meV}}$, ${{\rm RMSE}(l^{\rm Q}_{\rm max})=2.1\,{\rm meV}}$ 
\end{itemize}
As further explained in \change{the SI}, for \textit{really tight} defaults, the error introduced by non-regularized solvers for ill-conditioned matrices can reach the order of some $10^{-5}\,{\rm eV}$. Aiming for some $10^{-4}\,{\rm eV}$ accuracy, the regularized QR+SVD solver is then the safer choice. For the smaller basis sets, the error introduced by the solver is smaller and the error from the basis set is larger, so any solver can be used in that case.

For neutral solutes, the solvation free energies $\Delta G_{\rm solv}$ typically lie in the order of some tens to some hundreds of $\rm meV$.\cite{nonel} Tolerating even higher errors than for the suggested \textit{light} settings would thus introduce systematic errors in the order of magnitude of the quantity we want to determine and is therefore not advisable. Furthermore, the number and maximum value of outliers in \cref{fig:converge_lmax_rf,fig:converge_lmax_ep} increase dramatically beyond this point. This was also taken into consideration when choosing $l^{\rm Q}_{\rm max}=6$ for the \textit{tight} settings. Although the RMSE for $l^{\rm Q}_{\rm max}=4$ is closer to that for $l^{\rm R}_{\rm max}=8$, at $l^{\rm Q}_{\rm max}=6$ the number of systems with errors $>10\,{\rm meV}$ is significantly smaller, hinting at higher reliability.

\subsection{Model parameterization \label{sec:parameterization}}

Exactly analogously to the original MPE-1c case, the physical model parameters $\rho_{\rm iso}$ and $\alpha$ of MPE-$n$c need to be fitted to experimental reference data. Given that in the original MPE-1c method errors in the electrostatic contribution to the solvation free energy were likely compensated by non-electrostatic contributions, we cannot re-use the parameters computed for that model. Instead, we re-fit $\rho_{\rm iso}$ and $\alpha$ following the procedure outlined in 
earlier work.\cite{Sinstein} We apply the process mostly unchanged, simplifying a few technical details and sampling $\rho_{\rm iso}$ on a logarithmic grid. DFT calculations for all test molecules are correspondingly performed both in vacuum and in implicit solvent at isodensities $\rho_{\rm iso} = 10^x \frac{e}{\si{\angstrom}^3}$ with $x$ sampled on a grid from $-3$ to $-1$ in steps of $0.25$. The error function
\begin{equation}
    P(x,\alpha) = \sum_M^{\rm molecules}\left( \Delta G_{{\rm solv},M}^{\rm exp} - \left(\Delta G_{{\rm solv},M}^{\rm elstat}(x)+\alpha A_M(x) \right) \right)^2    
\end{equation}
is minimized, where $\Delta G_{{\rm solv},M}^{\rm elstat}(x)$ and $A_M(x)$ are cubic splines between the values of $x$ for which actual DFT calculations with $\rho_{\rm iso}(x)$ were performed.\cite{2020SciPy-NMeth}

All parameterization DFT calculations employed the \textit{really tight} settings of the MPE-$n$c solvation model. All other settings, such as the electronic basis sets and integration grids were set to the \textit{tight} defaults of \texttt{FHI-aims}\cite{Sinstein,aims-paper,aims-Havu2009,aims-Yu2018,aims-Ren2012,aims-Ihrig2015}. First, parameterization with the PBE\cite{PBE} functional was performed. The optimized isodensity value was found to lie in the upper part of the sampling grid. For subsequent parameterizations we thus only sampled the range from $-2$ to $-1$. In addition to PBE, parameters were optimized using the revPBE\cite{revpbe}, RPBE\cite{rpbe}, HSE06\cite{hse03,hse06}, BLYP\cite{Becke,LYP}, B3LYP\cite{B3LYP} and SCAN\cite{SCAN} functionals. As a consistency check, we also conducted a parameterization with PBE using MPE-1c with the original parameters $l_{\rm max}^{\rm R}=8$ and $l_{\rm max}^{\rm R}=6$.\cite{Sinstein} The parameterization results are shown in \cref{tab:params}.

\begin{table}[ht]
    \setlength{\tabcolsep}{0.5em}
    \centering
    \begin{tabular}{l|l|r|r|r|l}
        xc & solvent & $\frac{\rho_{\rm iso}}{1\,{\rm me \si{\angstrom}^{-3}}}$ &
        $\frac{\alpha}{1\,{\rm meV\si{\angstrom}^{-2}}}$ & $\frac{\rm RMSE}{1\,{\rm meV}} $ & $\frac{\rm RMSE}{1\,{\rm meV}}$ (p. HSE) \\
        \hline & & & & \\
        HSE06 & MPE-$n$c & \colorbox{lightgray}{33.68} & \colorbox{lightgray}{1.805} & 120 & \colorbox{lightgray}{120} \\
        PBE   & MPE-$n$c & 33.75 & 1.609 & 129 & \colorbox{lightgray}{132} \\
        revPBE& MPE-$n$c & 33.35 & 1.557 & 129 & \colorbox{lightgray}{134} \\ 
        RPBE  & MPE-$n$c & 33.11 & 1.527 & 133 & \colorbox{lightgray}{138} \\
        B3LYP & MPE-$n$c & 33.02 & 1.710 & 127 & \colorbox{lightgray}{127} \\
        BLYP  & MPE-$n$c & 32.74 & 1.513 & 137 & \colorbox{lightgray}{141} \\
        SCAN  & MPE-$n$c & 35.48 & 1.835 & 121 & \colorbox{lightgray}{123} \\
        PBE   & MPE-1c   & 33.36 & 1.281 & 125
    \end{tabular}
    \caption{Optimized parameters and RMSE of $\Delta G_{\rm solv}$ with respect to experimental references for water, using different solvent models and xc functionals. For the MPE-$n$c solvent model, the \textit{really tight} expansion orders from the previous section were used throughout. For MPE-1c, the originally published values\cite{Sinstein} of $l_{\rm max}^{\rm R}=8$ and $l_{\rm max}^{\rm R}=6$ were used. In the rightmost column, RMSEs calculated with the respective functional, but with $\alpha$ and $\rho_{\rm iso}$ optimized for HSE06 (cf.~highlighted values) are shown.}
    \label{tab:params}
\end{table}

The PBE family of xc functionals yields overall better results than the BLYP family, both for the generalized gradient approach (GGA) functionals (PBE, revPBE and RPBE vs. BLYP) and for the hybrid functionals (HSE06 vs. B3LYP). Within the PBE family of GGAs, the original PBE functional and revPBE yield a smaller RMSE than RPBE. The hybrid functionals generally yield an improvement of $\approx 10\,{\rm meV}$ compared to their GGA counterparts. The SCAN meta-GGA functional yields remarkably good agreement with experiment, almost matching the HSE06 hybrid functional and even outperforming B3LYP.

\begin{figure}[ht!]
\begin{subfigure}{.45\linewidth}
    \centering
    \includegraphics[width=\linewidth]{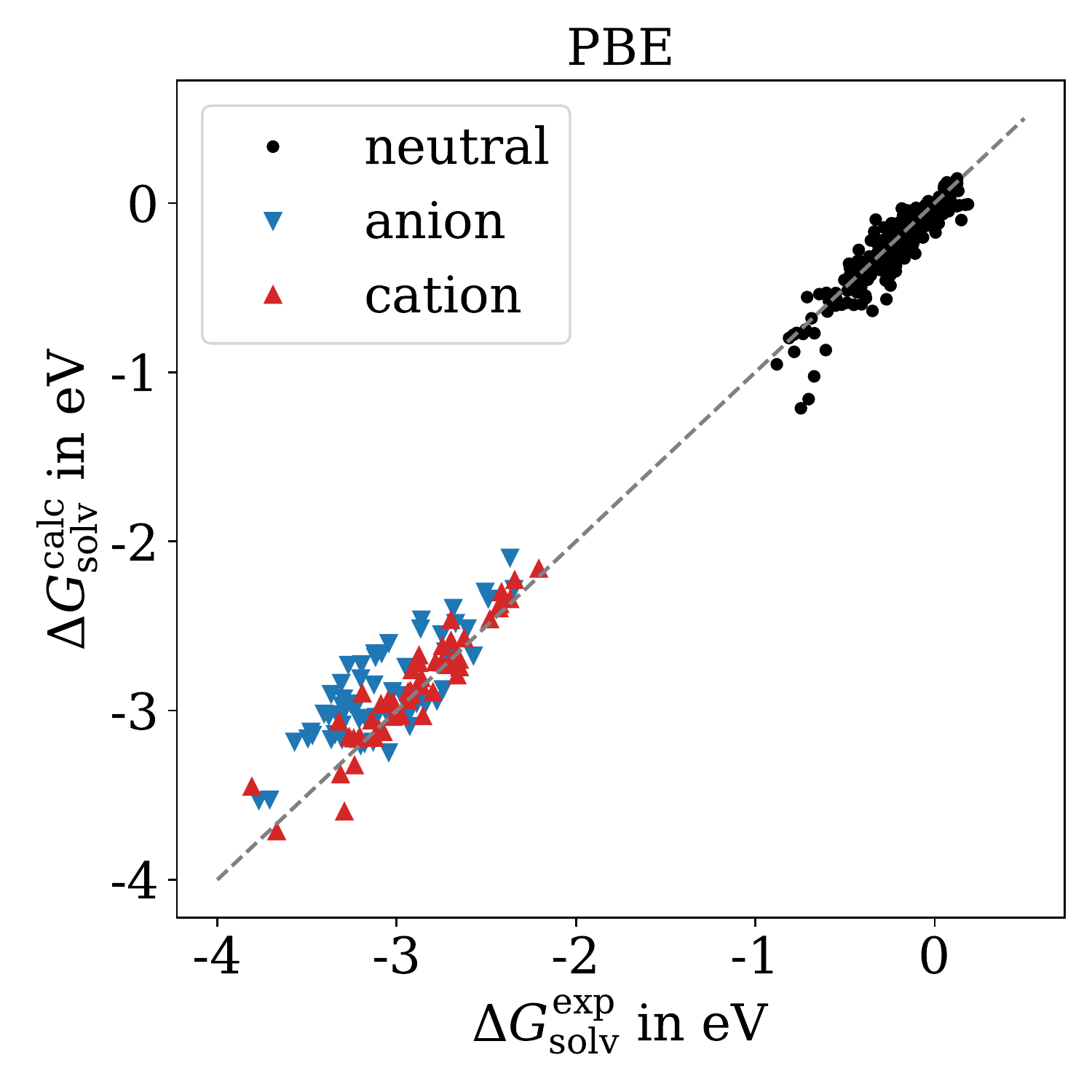}
\end{subfigure}
\begin{subfigure}{.45\linewidth}
    \centering
    \includegraphics[width=\linewidth]{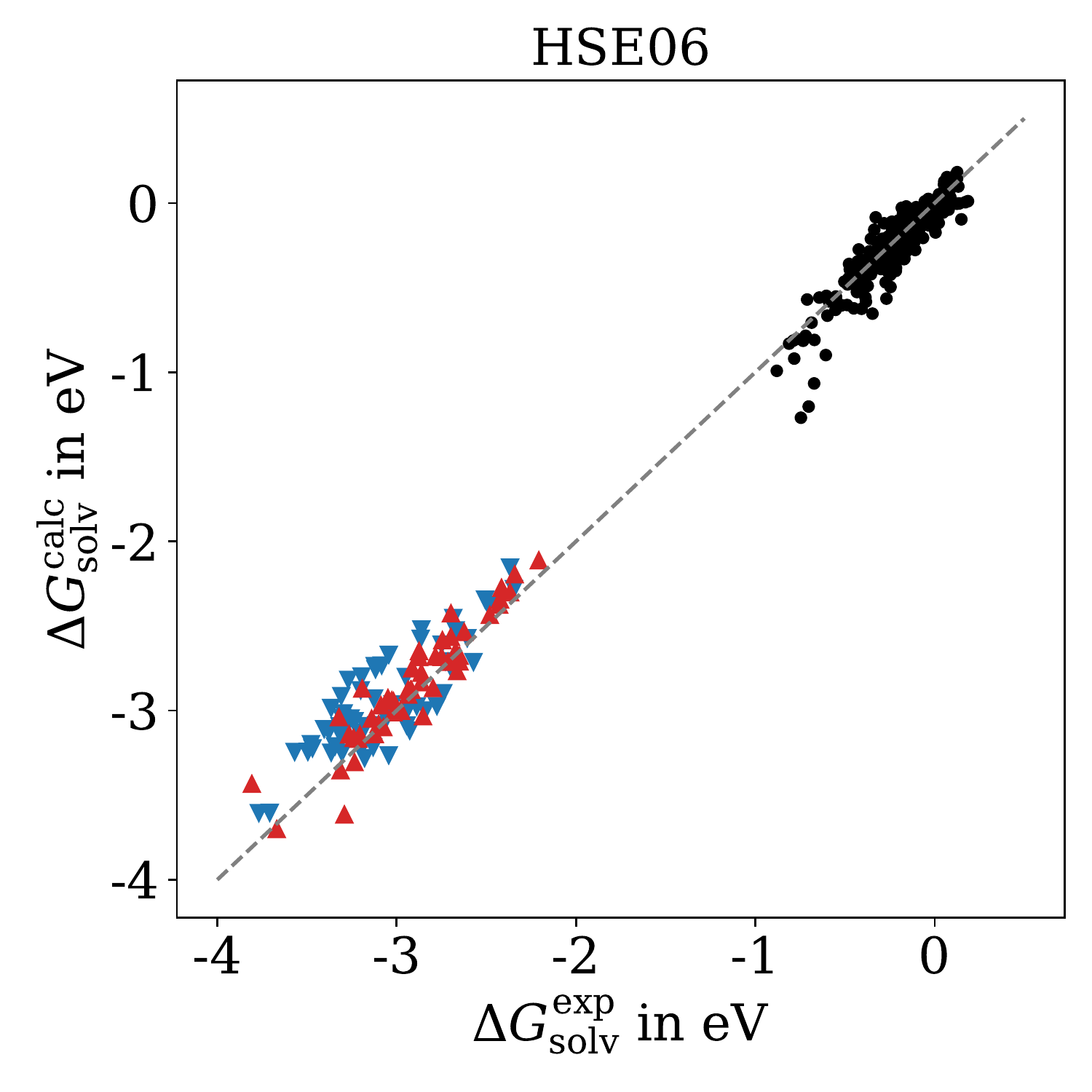}
\end{subfigure} \\
\begin{subfigure}{.45\linewidth}
    \centering
    \includegraphics[width=\linewidth]{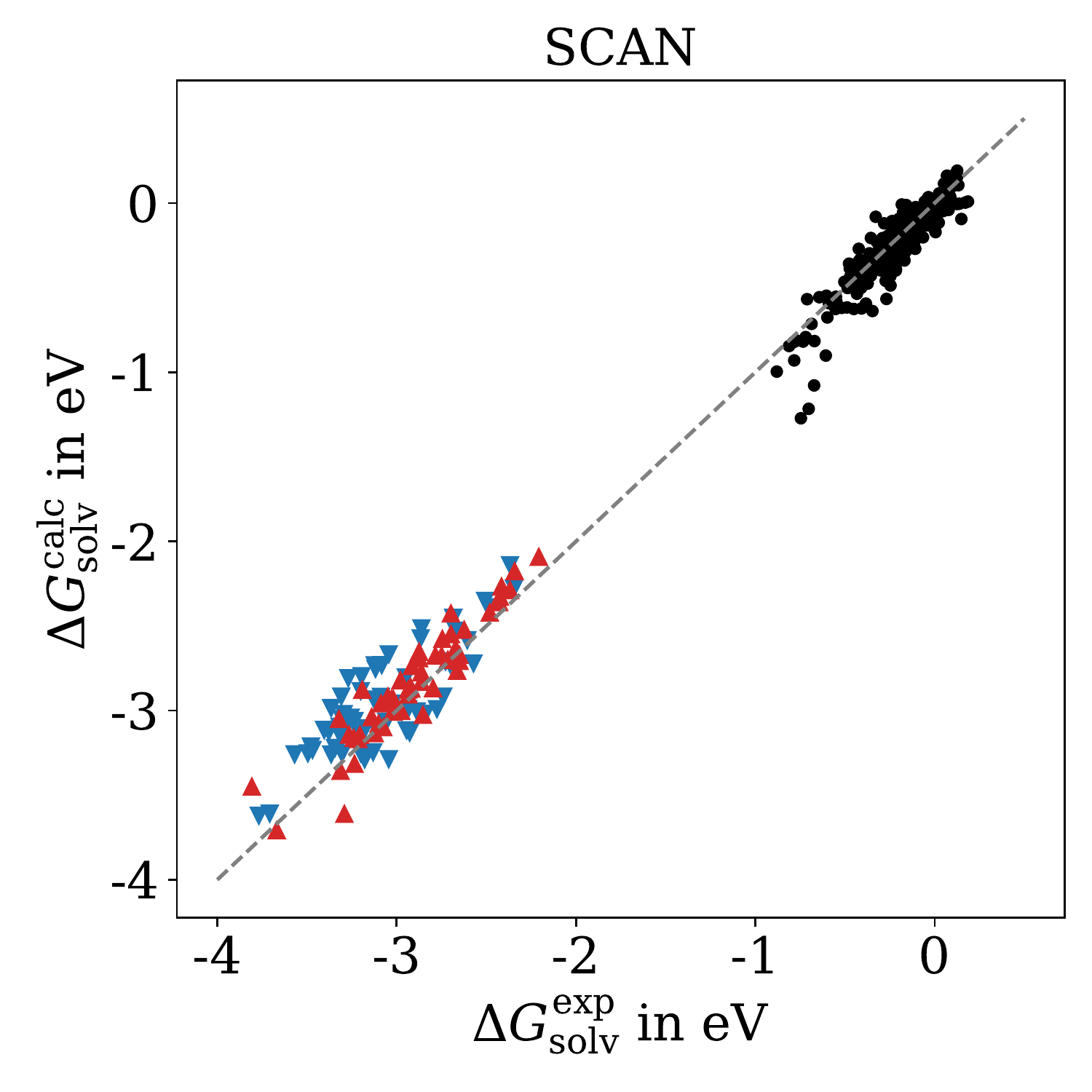}
\end{subfigure}
\begin{subfigure}{.45\linewidth}
    \centering
    \includegraphics[width=\linewidth]{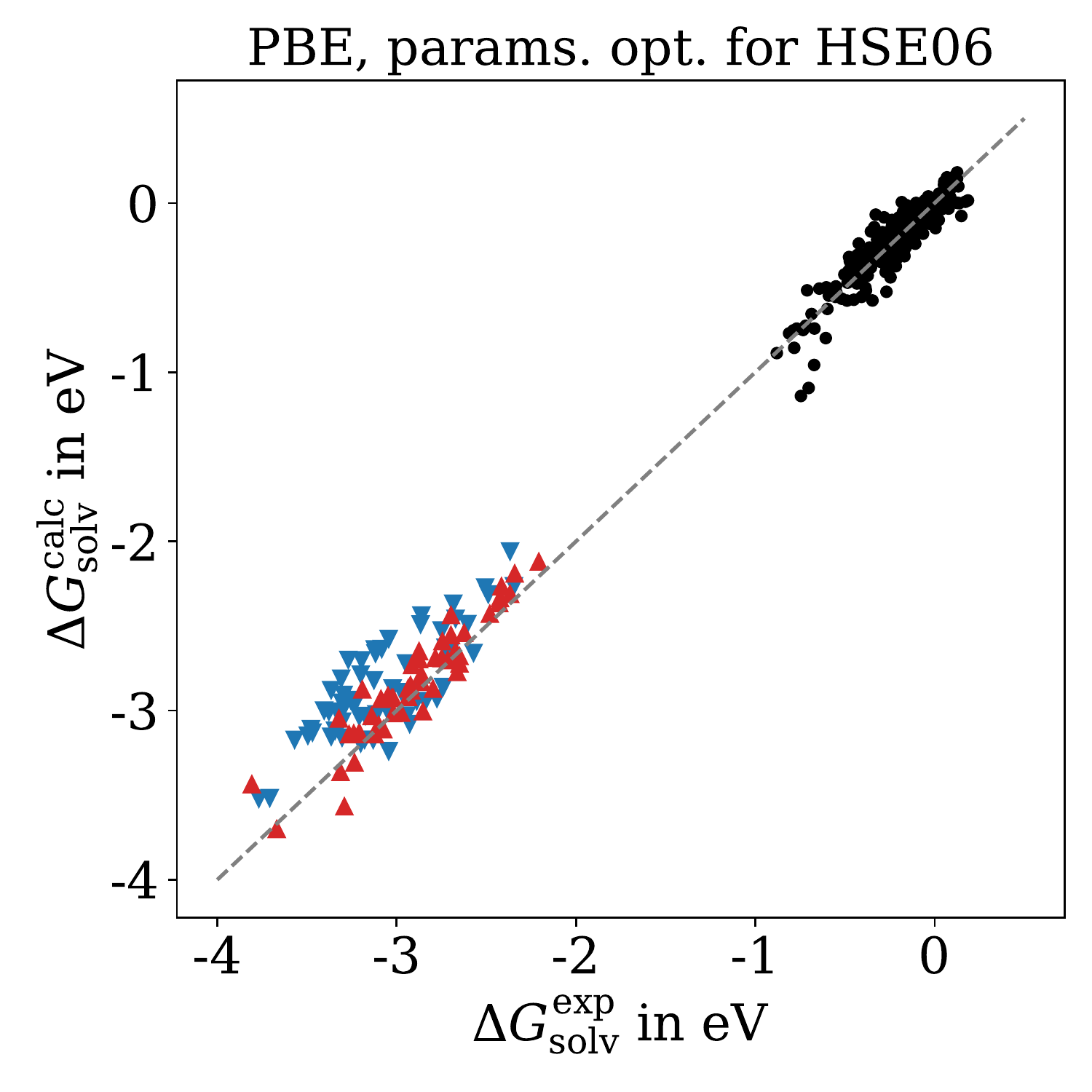}
\end{subfigure}
\caption{Calculated solvation free energies of solvation $\Delta G_{\rm solv}^{\rm calc}$ of our training set as calculated by MPE-$n$c plotted against experimental reference $\Delta G_{\rm solv}^{\rm exp}$, using 3 different xc functionals and with the optimized parameters from \cref{tab:params}. Grey dashed lines correspond to perfect agreement with experimental reference. The constituent terms of $\Delta G_{\rm solv}^{\rm calc}$, namely $\Delta G^{\rm elstat}_{\rm solv}$ and $\alpha A$ were calculated from cubic splines between the sampled values, see text. In the bottom right plot, the PBE functional was used together with parameters $\alpha$ and $\rho_{\rm iso}$ optimized for HSE06.
\label{fig:scatter}}
\end{figure}

In \cref{fig:scatter}, we show calculated $\Delta G_{\rm solv}$ of the solutes in the training set plotted against the experimental reference values for some selected functionals. It becomes clear that anionic solutes receive the largest improvement going from GGA to hybrid or meta-GGA functionals. A systematic underestimation (in terms of absolute values) of $\Delta G_{\rm solv}$, observed using GGA, is significantly improved by the hybrid and meta-GGA functionals, although it still persists to a lesser degree.

We also performed the parameterization using the alternative non-electrostatic model \cref{eq:nonel_Sinstein} that employs also the cavity volume. No significant improvement of our model's agreement with experiment compared to the simpler model \cref{eq:nonel_only_A} was obtained. For BLYP, it would for instance reduce the RMSE of $\Delta G _{\rm solv}$ by $2\,{\rm meV}$, and for all other xc functionals, the improvement is even smaller. We do thus not recommend usage of this alternative model and refrain from reporting the respective parameters to avoid confusion.

The parameters found for MPE-1c are in very good agreement with those found in earlier work\cite{Sinstein_thesis} using a slightly smaller training set. Remarkably, the RMSE of $\Delta G_{\rm solv}$ is approximately the same as with MPE-$n$c with the same xc functional, slightly smaller in fact. This reflects the aforediscussed combination of error cancellation and the smaller electrostatic error for the dominant smaller molecules in the test set. We have shown in \cref{sec:original_insuff} that the error in $\Delta G_{\rm solv}^{\rm elstat}$ is correlated with molecular size. Our descriptor for the non-electrostatic free energy contributions, $A$, is also correlated with molecular size by construction. A size-dependent systematic overestimation (on a signed scale) of $\Delta G_{\rm solv}^{\rm elstat}$ can thus be compensated for by a size-dependent systematic underestimation of $\Delta G^{\rm non-elstat}_{\rm solv}$. This can be seen in the optimized parameter $\alpha$ being much smaller for MPE-1c than for MPE-$n$c, leading to such an underestimation.

Interestingly, for MPE-$n$c smaller RMSEs seem to be correlated with higher $\alpha$, as shown in \cref{tab:params}. One possible interpretation is that some error cancellation of the above described kind still occurs with MPE-$n$c, presumably due to underestimation of the solute's polarization in the less exact functionals. Further research is needed to investigate this hypothesis.

We point out that the major part of the error cancellation in MPE-1c can be attributed to the choice of $l_{\rm max}^{\rm R}=8$ in our, as well as the original\cite{Sinstein,Sinstein_thesis} parameterization of MPE-1c. We have seen in \cref{sec:original_insuff} that this expansion order can be insufficient already for rather small solutes. In contrast, the number of solutes in the database\cite{MinSol} for which MPE-1c would not converge even at high expansion orders is small and makes only a minor contribution to the error cancellation in the original parameterization. However, this only concerns the parameterization process itself. In practical applications with large solutes, the problem of non-convergence in MPE-1c will emerge even when going to higher expansion orders.

This leads to the issue of transferability in MPE-1c. Consider a large solute. If we want to use parameters optimized at $l_{\rm max}^{\rm R}=8$, we do not know which $l_{\rm max}^{\rm R}$ to use for the large solute. On one hand, the systematic error cancellation at $l_{\rm max}^{\rm R}=8$ results from the parameterization process only for relatively small molecules, which make up the major part of the training data. The error in $\Delta G_{\rm solv}^{\rm non-elstat}$ depends linearly on the cavity surface area $A$. There is no reason to assume that the same holds strictly for $\Delta G_{\rm solv}^{\rm elstat}$. Therefore, the error cancellation is not guaranteed to work for large molecules at $l_{\rm max}^{\rm R}=8$. On the other hand, let us assume that $\Delta G_{\rm solv}^{\rm elstat}$ does converge with $l_{\rm max}^{\rm R}$---we have shown in \cref{sec:phyical_or_technical} that this can in some cases work even for large solutes. Simply picking a converged expansion order does not solve the problem then either. If $\Delta G_{\rm solv}^{\rm elstat}$ is converged, then $\Delta G_{\rm solv}^{\rm non-elstat}$ compensates for a non-existent error. One would have to guess at which expansion order the errors match in absolute value. The error cancellation is thus not transferable.

MPE-$n$c does not suffer from these transferability problems. Some error cancellation may still occur, as described above. However, this is due to inaccuracies in the xc functional, for which the assumption of linear correlation with system size is much more reasonable. As a safe choice, one may use the parameters optimized for HSE06, of which $\alpha$ is fairly repulsive, and apply them also when using other functionals. The resulting $\Delta G_{\rm solv}^{\rm elstat}$ will then be an upper estimate---on a signed scale---but not a strict upper boundary. In fact, this does not significantly impact agreement with experiment, as shown in the rightmost column of \cref{tab:params}. These parameters can thus be considered general parameters, as similarly suggested in earlier work.\cite{Sinstein} An exception is SCAN, for which the optimized $\alpha$ is actually larger than for HSE06. Here, the parameters optimized for SCAN itself should be at least equally as safe.

Despite these improvements, MPE-$n$c still calculates $\Delta G_{\rm solv}$ only to $\approx 100\,{\rm meV}$ accuracy. Possible sources of error are the simplistic model \cref{eq:nonel_only_A} for non-electrostatic solute-solvent interactions, which accounts for dispersive and repulsive interactions only in a statistical manner, the assumption of solute geometries which are rigid and also identical in vacuum and solution, and finally the lack of hydrogen bonds in our model. Generally lower errors for non-aqueous solvents in a similar model\cite{Hille} suggest that hydrogen bonds actually account for a considerable portion of the error, although the data sets used for these solvents in the cited study contained fewer solutes, which may also lead to lower errors.

\change{
\subsection{Computational performance \label{sec:performance}}

We test the computational performance of our method directly in the context in which it is implemented, as a feature of \aims. To have a state-of-the-art reference, we recompiled our method in a more recent version of \aims than the one used for the calculations shown in previous sections, namely version \texttt{211010}. All tests reported in the following were conducted on an \texttt{AMD Ryzen Threadripper 3970X 32-Core Processor}. The \texttt{GNU Fortran (Gentoo 10.3.0 p1) 10.3.0} compiler with flags \texttt{-O3 -march=native -fallow-argument-mismatch -ffree-line-length-none} was used for compilation. Performance relevant libraries used are \texttt{AMD optimized ScaLAPACK 3.0}\cite{ScaLAPACK-doc}, \texttt{AMD BLIS 3.0}\cite{BLIS1}, \texttt{AMD optimized libFLAME 3.0}\cite{libflame}, \texttt{ELPA 2020.05.001}\cite{aims-Marek2014,ELPA1} and \texttt{Open MPI v4.0.5}\cite{open_mpi}. We find for molecules in the size range considered here, that no considerable computational speedup is gained when increasing the number of cores beyond 16, as shown below. Thus, the scaling of computational cost with molecular size is tested using only 8 cores, to avoid artifacts due to inefficient parallelization. `Cores', in this section, refers to physical CPU cores.

In \cref{sec:parameterization}, we used the \textit{really tight} expansion orders from \cref{sec:convergence} to obtain as exact parameters as possible. For most practical applications, however, the smaller \textit{tight} expansion orders should be sufficient. These also allow for usage of the computationally more efficient QR or LSQR solvers, in contrast to the regularized QR+SVD solver necessary to deal with ill-conditioning at very high expansion orders. We test the scaling of computational time and memory requirement with solute size for the QR and LSQR solvers at \textit{tight} expansion orders, using the same test systems as in our convergence studies, cf.~\cref{sec:refdata}.

The scaling of computational time is compared to another commonly used implementation of an implicit solvation model, namely the SCCS model implemented in the Environ package\cite{nonel}, version 2.0. Already MPE-1c found agreement with experiment comparable to SCCS, making the latter an appropriate reference for such a performance test. For both models, mean absolute errors (MAE) around $50\,{\rm meV}$ for neutral (SCCS)\cite{nonel} or neutral and cationic (MPE-1c)\cite{Sinstein} solutes in water were reported, depending to some degree on the model parameters and specifics of the underlying DFT calculations. We used the electronic structure program \texttt{PWSCF v.7.0rc1} of the \textsc{Quantum ESPRESSO} package\cite{QE1,QE2,QE3} as a host for Environ. The \texttt{GNU Fortran (Gentoo 10.3.0 p1) 10.3.0} compiler with flags \texttt{-O3 -march=native -fallow-argument-mismatch} was used for compilation. Performance relevant libraries used are \texttt{AMD optimized ScaLAPACK 3.0}, \texttt{AMD BLIS 3.0}, \texttt{AMD optimized libFLAME 3.0} and \texttt{Open MPI v4.0.5}. The \texttt{H.pbe-rrkjus\_psl.1.0.0.UPF} (H) and\\ \texttt{x.pbe-n-rrkjus\_psl.1.0.0.UPF} (all other elements) ultrasoft pseudopotentials (USPP) from \texttt{http://www.quantum-espresso.org} were used. Energy cutoffs for charge density and wavefunctions were chosen as suggested in the pseudopotential files. The parabolic point-counter-charge (PCC) correction scheme\cite{PCC,PCC_erratum} implemented in Environ was used to remove interactions with periodic images in all directions, both in vacuum and implicit solvent calculations. The PCC correction, in its current implementation, requires cubic simulation cells. Cell sizes of $a=1.1\,d_{\rm max} + 5\, \si{\angstrom}$ were used, with $d_{\rm max}=\max(d_x,d_y,d_z)$, where $d_{x/y/z}$ is the extent of the solute in $x/y/z$ direction. These extents were calculated taking into account the van-der-Waals radii of the atoms, as obtained from the Atomic Simulation Environment ASE\cite{ASE}. To keep the simulation cells small, prior to determining the cell size the solute molecules were rotated such that the axis corresponding to the lowest principal moment of inertia was oriented along the space diagonal. Environ's `full' cavity definition was used. The same randomized $\varepsilon$ as for the MPE calculations were used. The randomized $\rho_{\rm iso}$ were converted to the density thresholds $\rho_{\rm min}, \rho_{\rm max}$ of SCCS using the transformation formula from ref.~\citenum{Hille}, with the generic smoothness parameter $\delta_n=2.0$ reported therein. Pressure, surface tension and electrolyte concentration were all set to 0.

Our method is implemented for parallel execution. The scaling with physical cores is tested on the largest of the test systems (entry number test1012 in the Minnesota Solvation Database\cite{MinSol}), containing 28 non-H atoms.

\begin{figure}[!ht]
\begin{subfigure}{.45\linewidth}
    \centering
    \includegraphics[width=\linewidth]{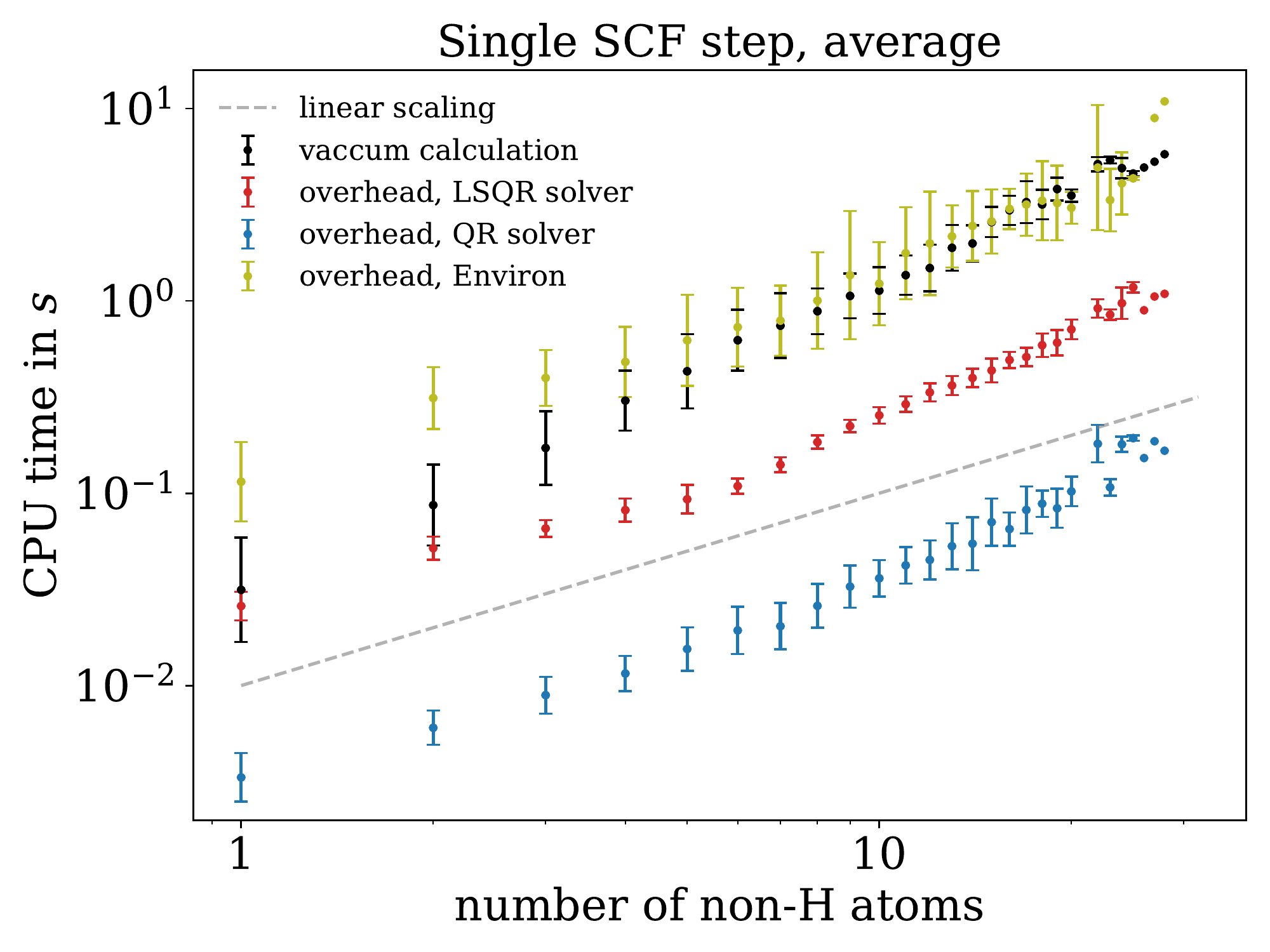}
\end{subfigure}
\begin{subfigure}{.45\linewidth}
    \centering
    \includegraphics[width=\linewidth]{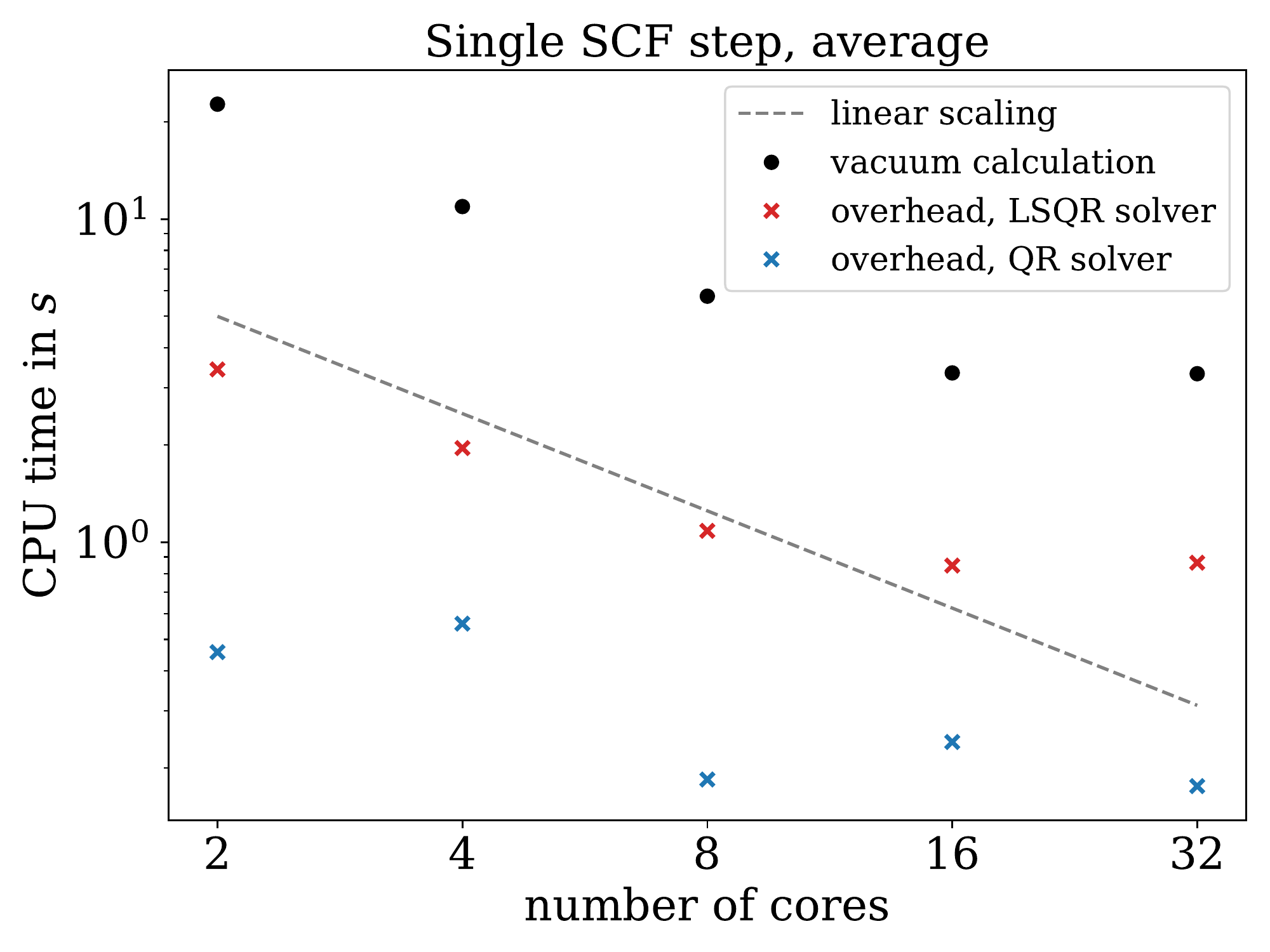}
\end{subfigure} \\
\begin{subfigure}{.45\linewidth}
    \centering
    \includegraphics[width=\linewidth]{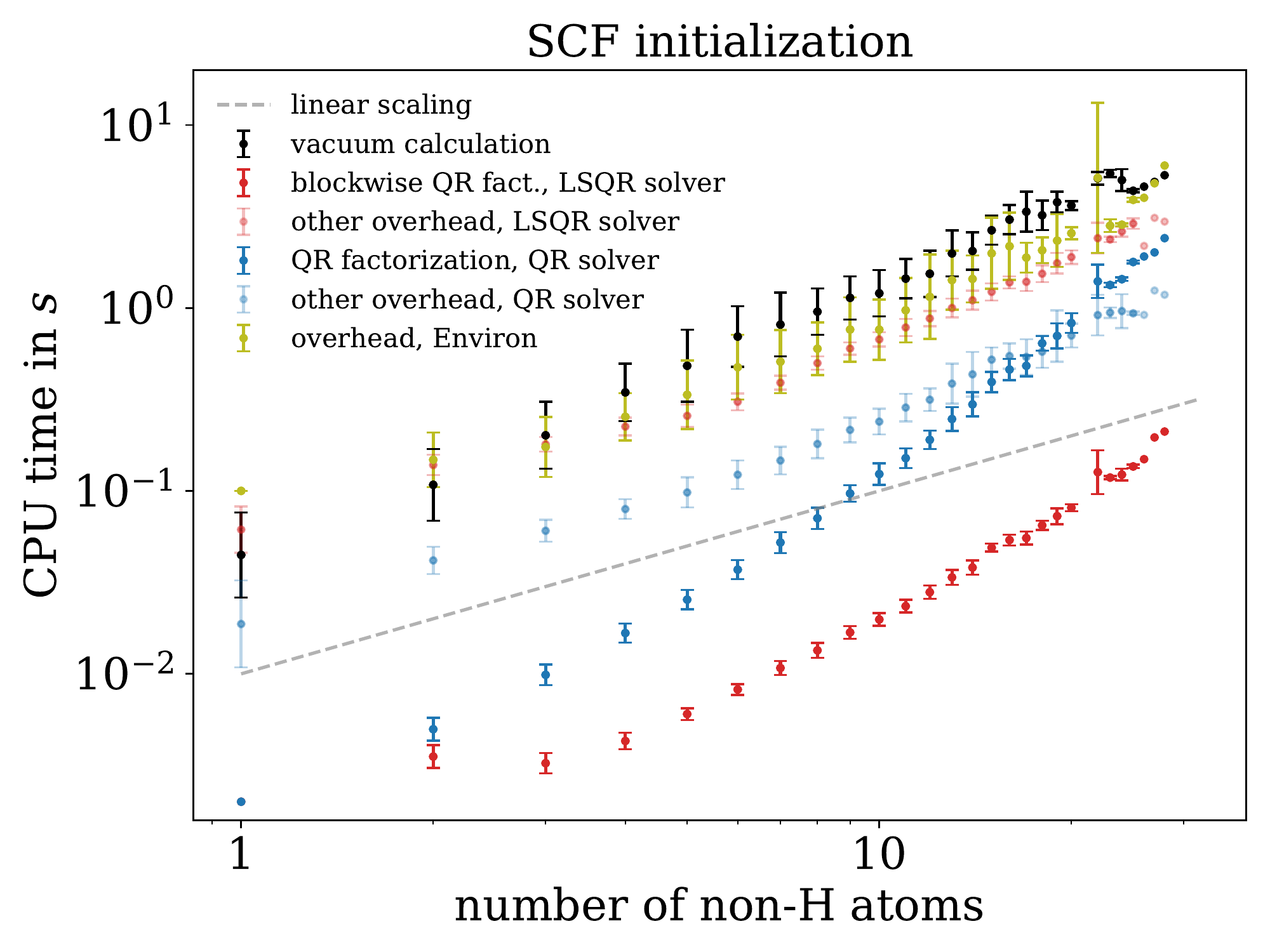}
\end{subfigure}
\begin{subfigure}{.45\linewidth}
    \centering
    \includegraphics[width=\linewidth]{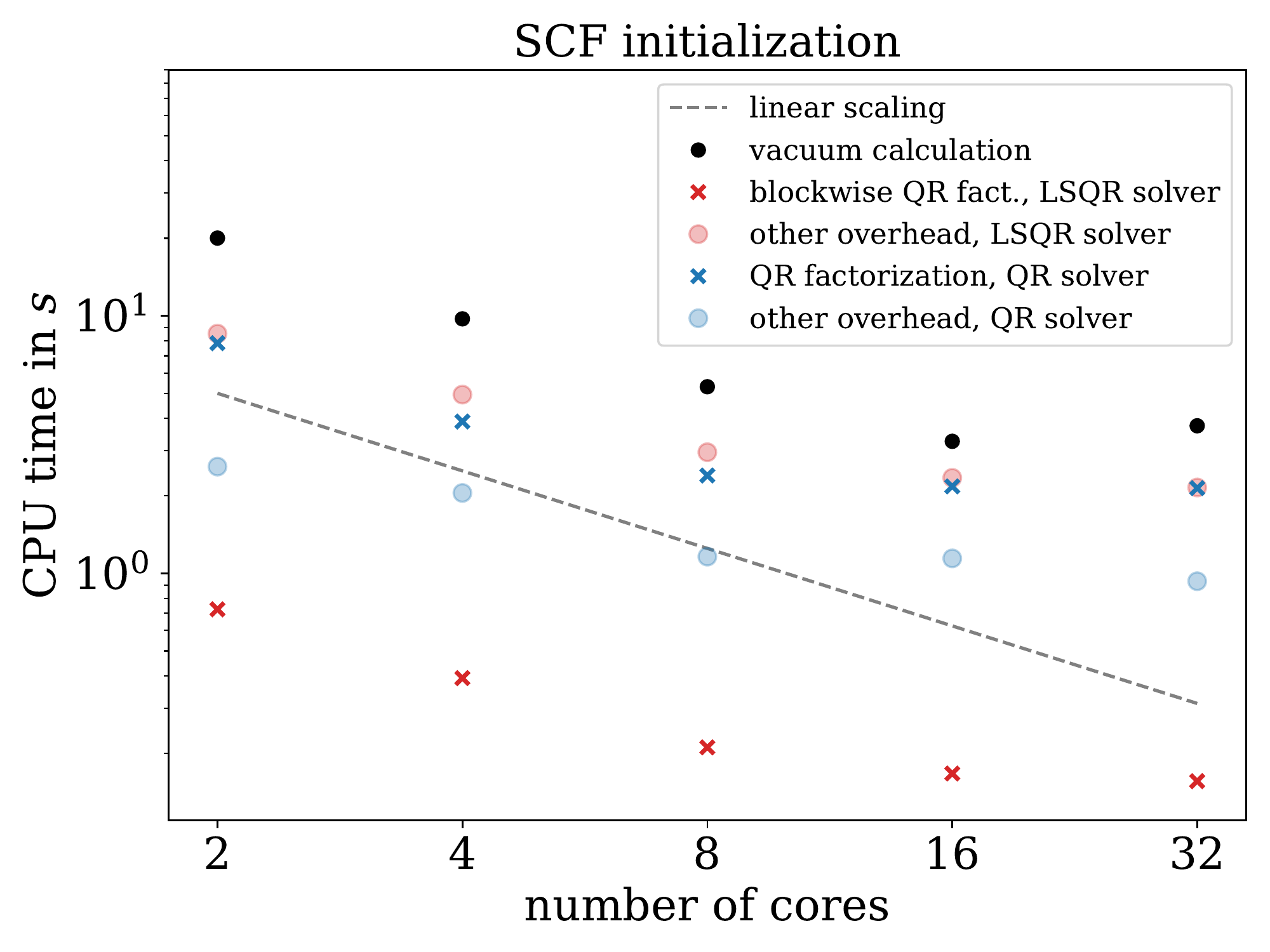}
\end{subfigure}
\caption{\change{Scaling of CPU time with solute size on 8 cores (left) and with number of cores for largest solute in test set (right), all on a double logarithmic scale. Hypothetical linear scaling shown at arbitrary offset for reference. Negative outliers excluded from analysis. Overhead calculated as difference between values in implicit solvent and vacuum calculation. Left: Mean over solutes of same size shown as dot, root mean square deviation shown as error bar, both calculated in logarithmic space. Overhead of Environ\cite{nonel} in a \textsc{Quantum ESPRESSO}\cite{QE1,QE2,QE3} calculation shown for reference. Timing for \textsc{Quantum ESPRESSO} without implicit solvation not shown. Top: Time for 1 SCF step, averaged over all SCF steps except the last, which is typically shorter. Bottom: Time for SCF initialization. `Other overhead' calculated as overhead minus time for (blockwise) QR factorization.}
\label{fig:timing}}
\end{figure}

\begin{figure}[!ht]
\begin{subfigure}{.45\linewidth}
    \centering
    \includegraphics[width=\linewidth]{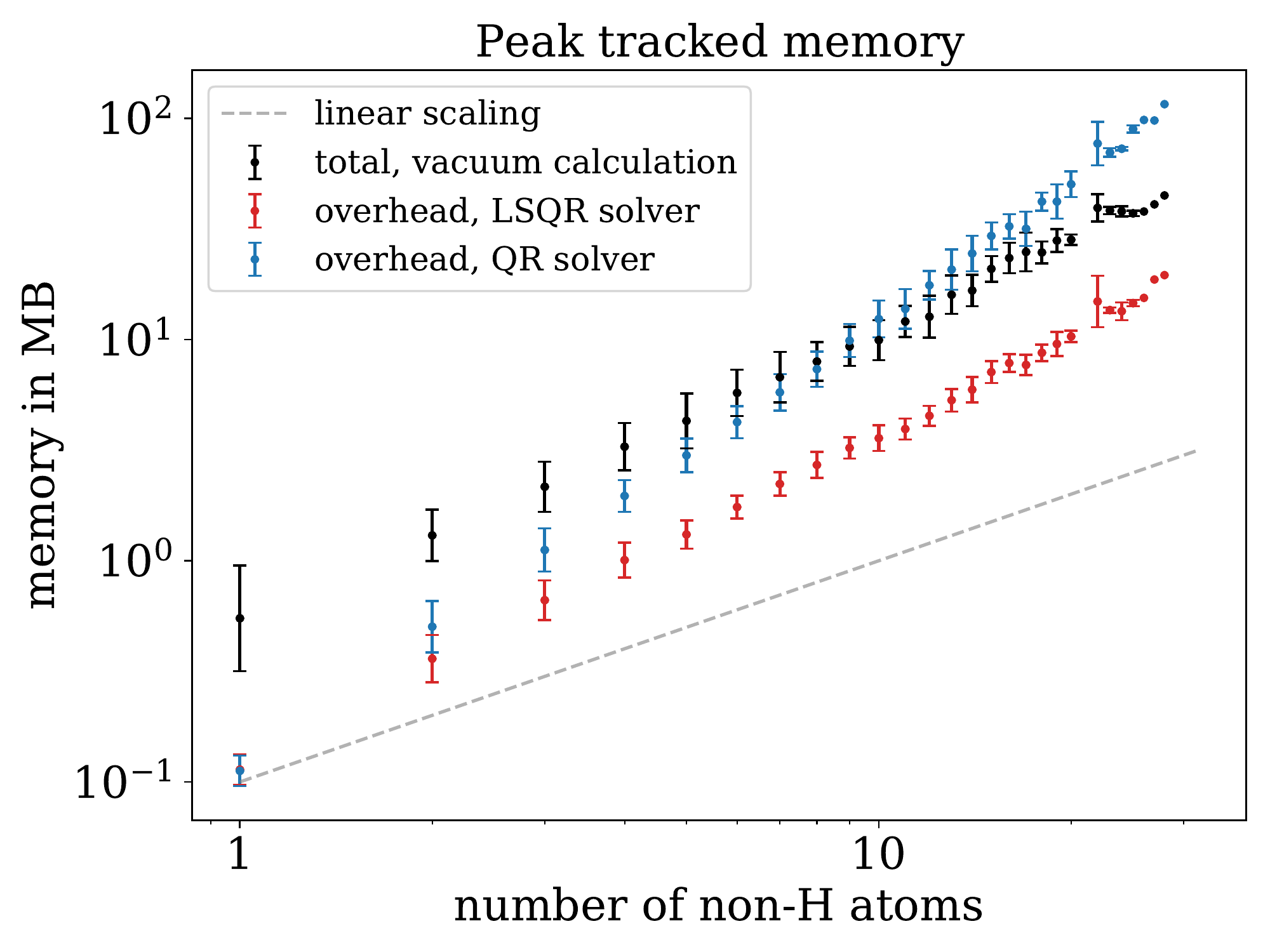}
\end{subfigure}
\begin{subfigure}{.45\linewidth}
    \centering
    \includegraphics[width=\linewidth]{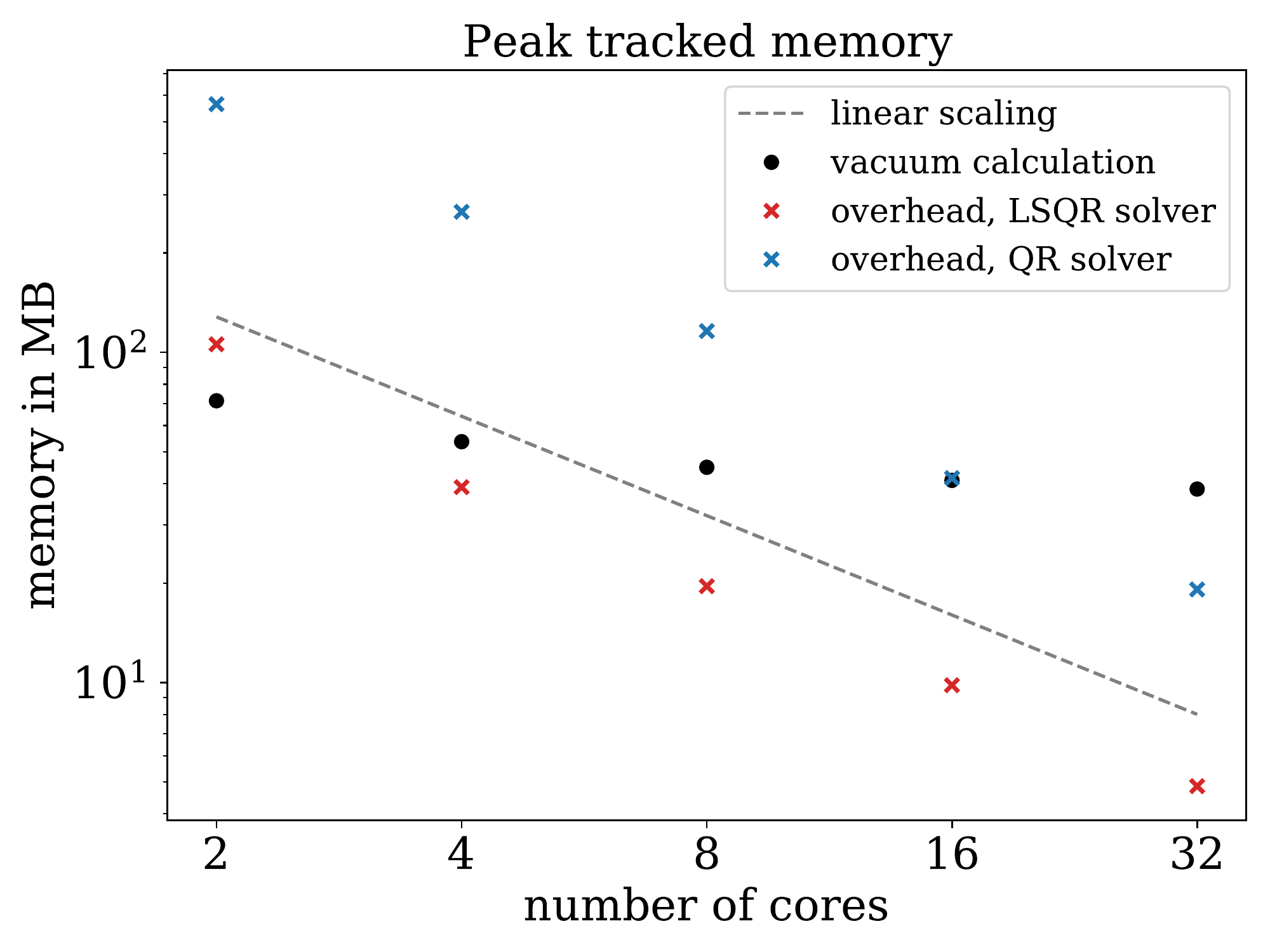}
\end{subfigure}
\caption{\change{Scaling of highest tracked memory across all processes with solute size on 8 cores (left) and with number of cores for largest solute in test set (right), all on a double logarithmic scale. \aims does not track small arrays and scalars; real memory usage will be slightly larger.}
\label{fig:memory}}
\end{figure}

The main results of these tests are reported in \cref{fig:timing,fig:memory}. While the LSQR solver takes consistently more time for one SCF step than the QR solver, it is never the bottleneck of the calculation in the sampled size range, except for very small solutes. Both solver types seem to scale almost linearly with system size, making the overhead of our method in one SCF step generally unproblematic. In the SCF initialization, both solver types cause a significant overhead. Typically, the initialization is not the bottleneck of a calculation in \aims, so the overhead here is negligible as long as it does not exceed the initialization time of a vacuum calculation by orders of magnitude. In the sampled size range, this is never the case. However, it is clear that the scaling for the QR factorization in the QR solver is unfavorable, and this step will eventually become a bottleneck for larger solutes. The scaling of the blockwise QR factorization in the LSQR solver is not linear either, but it is generally much faster than for the QR solver, making LSQR more suitable for very large systems. Overhead other than the factorizations seems to scale no worse than the encompassing DFT initialization in both cases, with a higher offset for LSQR. Similar observations can be made in the memory requirements. Again, the scaling of the QR solver with system size is unfavorable, approximately quadratic in this case. For the LSQR solver, it is, again, much flatter, albeit not perfectly linear.

Compared to Environ\cite{nonel}, MPE-$n$c with both solvers is consistently faster in the SCF steps, within the sampled size range. In the SCF initialization, the overhead is comparable in the sampled size range. This comparison is, to some degree, biased by the different design choices in the SCCS and MPE-$n$c models, as well as the characteristics of the electronic structure software used as host. Most importantly, \textsc{Quantum ESPRESSO}\cite{QE1,QE2,QE3} is a plane wave code, formally requiring periodic boundary conditions (PBC) which are then compensated for e.g. by the PCC\cite{PCC,PCC_erratum} scheme for non-periodic systems. The necessity of PBC can increase computational costs if significant portions of the simulation cell are empty. \aims on the other hand uses atom-centered basis functions, allowing it to conduct the computation entirely in real space. This difference also affects the respective implicit solvation methods. Despite these limitations, this comparison illustrates the computational efficiency of MPE-$n$c, not only in the given context of \aims.

In terms of parallelization over cores, both solvers scale fairly well. Within one SCF step, the LSQR solver scales approximately equally as well as the encompassing DFT calculation. No considerable speedup is gained for the test system when going beyond 16 cores. This is the case both for the MPE method and the vacuum calculation. For larger solutes, parallelization over a larger number of cores will likely become more efficient. For the QR solver, no clear scaling is observed here, but the overhead is already negligibly small. In the SCF initialization, all parts of our method scale well within a certain range. Again, both for the encompassing DFT calculation and the MPE method, no speedup is gained when going beyond 16 cores, although MPE appears to reach a plateau already at slightly fewer cores. Nonetheless, all parts of both solvers consistently stay below the time for the DFT initialization. In terms of memory, the distribution is trivial and scales linearly with the number of cores for both solvers.

Finally, we briefly address computational scaling from a more theoretical point of view. The size of the central matrix $\mathbf{A}$ in \cref{eq:SLE} scales $O(N^2)$ with system size $N$. Each non-H atom adds a constant number $(l_{\rm max}^{\rm R}+1)^2+(l_{\rm max}^{\rm Q}+1)^2$ of columns to $\mathbf{A}$. The (target) number of rows depends linearly on the number of columns via $d_{\rm det}$. The computational time of the QR factorization will thus become the bottleneck at some point. The LSQR solver largely circumvents this issue by exploiting sparsity. In the $(l_{\rm max}^{\rm R}+1)^2$ columns corresponding to the basis set of one subcavity, only approximately $2d_{\rm det} (l_{\rm max}^{\rm R}+1)^2$ rows are filled. Each non-H atom thus only adds an approximately constant number of non-zero matrix elements in these blocks. Nonetheless, the matrix block corresponding to the basis set for $\Phi_{\rm Q}$ does not get sparser in the present method. Each non-H atom adds $(l_{\rm max}^{\rm Q}+1)^2$ columns, which have filled rows for all cavity surface points, the number of which also grows approximately linearly with system size. Thus, the LSQR solver will approach at least $O(N^2)$ scaling in the limit of very large solutes, as well. However, it does so at a significantly lower offset than the QR solver. The issue of choosing the basis set for $\Phi_{\rm Q}$ in a way that will sparsify $\mathbf{A}$ is beyond the scope of the present paper and left to future work.

}

\section{Conclusions \change{and outlook}}

We have shown that MPE-1c with the standard $l_{\rm max}^{\rm R}=8$ leads to a systematic underestimation of the electrostatic solute-solvent interaction. Already in the original publication\cite{Sinstein} it was noted that potentially much higher expansion orders would be necessary for large solutes. We have shown, however, that the issue can sometimes occur already in relatively small molecules. Furthermore, we have seen that for larger solutes, especially using high $\rho_{\rm iso}$, increasing the expansion order is not always a viable solution, with the multipole series converging slowly or not at all.

As a remedy for this issue, we have modified the MPE method, termed the MPE-$n$c method, by separating the solvent cavity into small subcavities, centered around the solute's non-H nuclei. The dielectric response of the solvent is expanded in an individual multipole series in each of the subcavities. In practice, this ensures a fast convergence of the multipole series. We have parameterized an implicit solvation model for water using our modified electrostatic MPE-$n$c approach and a simplistic model for non-electrostatic free energy contributions. Comparing our results to an equivalent parameterization using MPE-1c reveals a systematic error cancellation between electrostatic and non-electrostatic model terms in MPE-1c. While yielding surprisingly accurate results, this error cancellation can not be expected to hold for larger, more complex solutes, limiting the original method's transferability. These problems are not present in the MPE-$n$c approach which allows for a much more reliable reproduction of the dielectric response of a polarizable medium.

\change{The practical applicability of our method is currently limited by the lack of atomic forces. While these lie beyond the scope of the present, self-contained work, their derivation and implementation are an obvious next step which we plan to address in future work.}

\begin{acknowledgement}
The authors thank Markus Sinstein and Sebastian Matera for various insightful discussions during the development of our modification of their original method. Furthermore, we thank Konstantin Jakob, who conducted exploratory studies on a possible alternative parameterization procedure and alternative models for the non-electrostatic free energy contributions. The authors also acknowledge financial support from the ``Solar Technologies go Hybrid'' initiative of the state of Bavaria.
\end{acknowledgement}

\appendix

\change{

\section{Supporting Information}

Supporting information (SI) is available for this article. There, we describe the technical details of the central linear algebra problem of our method, as well as the boundary discretization algorithm. Lastly, we provide a minimal example for the user input required for our method.

This information is available free of charge via the Internet at http://pubs.acs.org

}
\bibliography{refs}

\end{document}